\documentclass[usenatbib,babel]{mn2e}

\usepackage[english,english]{babel}
\usepackage{amsmath}
\usepackage{amssymb,amsfonts,textcomp}
\usepackage{array}
\usepackage{supertabular}
\usepackage{hhline}
\usepackage{hyperref}
\usepackage[usenames]{color}

\def\gtrsim{\lower.5ex\hbox{$\; \buildrel > \over \sim \;$}}
\usepackage{graphicx}

\definecolor{grey}{rgb}{0.75,0.75,0.75}
\definecolor{Orange}{rgb}{1.0,0.5,0.15}
\definecolor{brown}{rgb}{0.7,0.25,0.0}
\definecolor{pink}{rgb}{1.0,0.5,0.5}
\definecolor{darkerred}{rgb}{0.8,0,0}
\definecolor{darkerblue}{rgb}{0,0,0.8}
\definecolor{Blue}{rgb}{0,0.08,0.65}
\definecolor{Red}{rgb}{0.65,0.08,0.05}
\definecolor{Green}{rgb}{0.15,0.45,0.25}

\begin{document}

\author[C. Welker et al.]{
\parbox[t]{\textwidth}{C. Welker$^{1,2,4,5}$\thanks{E-mail: welkerc@mcmaster.ca}, J. Bland-Hawthorn$^{3,4}$, J. Van de Sande$^{3,4}$, C. Lagos$^{1,4}$, P. Elahi$^{1,4}$, D. Obreschkow$^{1,4}$, J. Bryant$^{3,4}$, C. Pichon$^{5,12}$, L. Cortese$^{2,4}$, S. N. Richards$^{6}$, S. M. Croom$^{3,4}$, M. Goodwin$^{7}$, J. S. Lawrence$^{8}$, S. Sweet$^{9,4}$, A. Lopez-Sanchez$^{8}$, A. Medling$^{10,11}$, M. S. Owers$^{13}$, Y. Dubois$^{5}$, J. Devriendt$^{14,15}$}
\vspace*{6pt} \\ 
$^{1}$ Department of Physics and Astronomy, McMaster University, Hamilton, Ontario, Canada\\
$^{2}$ ICRAR, University of Western Australia, Crawley, Perth, 6009, Western Australia\\
$^{3}$ Sydney Institute for Astronomy, School of Physics, A28, The University of Sydney, NSW, 2006, Australia\\
$^{4}$ ARC Centre of Excellence for Astrophysics in 3 Dimensions (ASTRO 3D)\\
$^{5}$ CNRS and UPMC Univ. Paris 06, UMR 7095, Institut d'Astrophysique de Paris, 98 bis Boulevard Arago, F-75014 Paris, France\\
$^{6}$ SOFIA Science Center, USRA, NASA Ames Research Center, Building N232, MS 232-12, P.O. Box 1, Moffett Field, CA 94035-0001, USA \\
$^{7}$ Australian Astronomical Observatory, 105 Delhi Rd, North Ryde, NSW 2113, Australia \\
$^{8}$ Lawrence Australian Astronomical Optics, Macquarie, Macquarie University, NSW 2109, Australia \\
$^{9}$ Centre for Astrophysics and Supercomputing, Swinburne University of Technology, PO Box 218, Hawthorn, VIC 3122 \\
$^{10}$ Ritter Astrophysical Research Center University of Toledo Toledo, OH 43606, USA  \\
$^{11}$ Research School for Astronomy and Astrophysics Australian National University Canberra, ACT 2611, Australia\\
$^{12}$ Korea Institute of Advanced Studies (KIAS) 85 Hoegiro, Dongdaemun-gu, Seoul, 02455, Republic of Korea\\
$^{13}$ Department of Physics and Astronomy, Macquarie University, NSW 2109, Australia\\
$^{14}$ Sub-department of Astrophysics, University of Oxford, Keble Road, Oxford, OX1 3RH, United Kingdom\\
$^{15}$ Observatoire de Lyon, UMR 5574, 9 avenue Charles Andre, Saint Genis Laval 69561, France\\
}
\date{Accepted . Received ; in original form }

\title[]{The SAMI Galaxy Survey: First detection of a transition in spin orientation with respect to cosmic filaments in the stellar kinematics of galaxies.}

\maketitle

\begin{abstract}
{We present the first detection of mass dependent galactic spin alignments with local cosmic filaments with $>2\sigma$ confidence using IFS kinematics. The 3D network of cosmic filaments is reconstructed on Mpc scales across GAMA fields using the cosmic web extractor DisPerSe. We assign field galaxies from the SAMI survey to their nearest filament segment in 3D and estimate the degree of alignment between SAMI galaxies' kinematic spin axis and their nearest filament in projection.  Low-mass galaxies align their spin with their nearest filament while higher mass counterparts are more likely to display an orthogonal orientation. The stellar transition mass from the first trend to the second is bracketed between $10^{10.4}\, M_{\odot}$ and $10^{10.9}\, M_{\odot}$, with hints of an increase with filament scale. Consistent signals are found in the Horizon-AGN cosmological hydrodynamic simulation. This supports a scenario of early angular momentum build-up in vorticity rich quadrants around filaments at low stellar mass followed by progressive flip of spins orthogonal to the cosmic filaments through mergers at high stellar mass. Conversely, we show that dark-matter only simulations post-processed with a semi-analytic model treatment of galaxy formation struggles to reproduce this alignment signal. This suggests that gas physics is key in enhancing the galaxy-filament alignment.}
\end{abstract}


\section{Introduction}

In standard cosmology, the highly anisotropic distribution of matter on scales of 10 to 100 Mpc, referred to as the cosmic web, is made of massive clusters (where brightest galaxies form and reside with their satellites) connected through dense filaments - along which smaller galaxies drift - framing the honeycomb-like structure of walls of lower density. Such structure naturally arises from the anisotropic gravitational collapse of an initially Gaussian random field of density perturbations \citep{Zeldo70,Shandarin89,Peebles80,Bond96}. Haloes form and reside within the overdensities of the cosmic web, accreting smooth material and smaller haloes via filaments they contributed to form in between them \cite[see][for details]{Bond96}. Knots at the intersection of several contrasted filaments house clusters, the largest virialized objects in the Universe. On such scales, the filamentary pattern of the cosmic web is apparent in all large-scale galaxy surveys \cite[e.g.][]{Delapparent86,Colless03a,Doroshkevich04,Alpaslan14a}, traced by the galaxy distribution.

This structure of matter on large scales conditions the geometry and dynamics of gas and galaxy flows from early times onwards. As galaxies migrate from voids towards walls, from walls towards filaments and eventually from filaments towards nodes - most of them are found in the immediate vicinity of filaments - the galaxy distribution is affected, leading among other things to anisotropic infall of satellites into massive haloes \citep{Aubert04,Knebe04,Wang05,Zentner05,Yang06} and more generally to preferential orbits for mergers, seemingly impacting the properties of the remnants \citep{Holmberg69,Zaritsky97,Brainerd05,Yang06,Sales09,Wang10,Nierenberg12, smithetal15,Huang17}. At higher redshifts ($z>1.5$), simulations predict that gas trapped in collapsed dark matter halos is funnelled through cold filamentary streams shaped by the cosmic filaments and advected towards the centre of forming galaxies to which it transfers part of its angular momentum \citep{Birnboim03,Ocvirk08,Dekel09,Pichon11,Danovich12}. Over cosmic time, gas rarefies and the environment evolves through small-scale processes including hydrodynamic instabilities or feedback from black holes and supernovae. These may drastically affect the morphology and internal dynamics of galaxies and in part damp initial correlations with the cosmic web. 

An important predicted effect of the cosmic web on growing haloes and galaxies was first identified in N-body cosmological simulations \citep{Aubert04,Aragon07,Hahn07,Paz08,Bett12,Codis12} and later extended to galaxies in hydrodynamic runs \citep{Hahn07,Dubois14,Codis18}. It is the mass-dependent trend of galactic spins to align with nearby filaments, which reveals a strong connection between the complex accretion of material on galactic scales and cosmic flows on extra-galactic scales. These studies find that low-mass galaxies tend to align their spin with their nearby filaments while their most massive counterparts have their spin orthogonal to the filament. Theoretical works extending tidal torque theory \citep[see][for a detailed introduction and references]{Porciani02} to constrained anisotropic environments \citep{Codis15} also predict a similar transition of the spin orientation as haloes as they drift along the filaments and grow in mass. Hints of such a transition for galaxies have also been identified in the SDSS using the projected shape as a proxy for galaxy spin by \cite{Tempel13a,Tempel13b,Pahwa16} and \cite{Chen19} but with a limited significance and distinct probabilistic filament reconstruction methods. Since its detection in simulations, this faint spin flip signal has therefore remained elusive in observations. Indeed, its detection is made very difficult by the lack of large samples of galaxies with kinematic measurements and the use of filament reconstruction methods not focused on recovering precise orientations for filaments across scales but rather on identifying entire patches of a survey as a certain type of cosmic feature (namely filament, wall, node, void) \citep[see][for references]{Libeskind18}.

Over the last decade, robust methods of cosmic web features identification have been developed \citep{Sousbie11,Cautun13,Felix15}, with a stronger focus on recovering the multiscale nature and directionality of filaments over large scales and are now being applied on spectroscopic and photometric surveys such as COSMOS allowing cutting-edge analysis of this interplay on real datasets \citep{Laigle18,Malavasi17,Kraljic18}. These studies join several others \citep{Rojas04, Guo15, Beygu16, Kleiner17,Kuutma17} which investigated a degree of segregation of galaxy properties such as star formation rate, morphology, colours or even atomic hydrogen content across the various features of the cosmic web, distinct from the simple effect of mass and local density. Note however that such effects remain faint compared to mass and local density driven processes \citep{Eardley15, Alpaslan15, Alpaslan16}, hence the necessity of very large samples and very large volumes to detect it.

Unlike the above mentioned scalar properties, anisotropies in spin orientations or kinematical properties with respect to the cosmic web are in theory independent of purely mass or density driven effects but are harder to obtain with sufficient accuracy. While still restricted in terms of statistics, Integral Field Spectroscopy (IFS) surveys such as SAMI (Anglo-Australian Observatory) \citep{Croom12} or MaNGa \citep{Bundy15} provide high quality stellar and gas kinematics within approximately one effective radius across a wide range of environments and for an increasing number of low-redshift galaxies with near kiloparsec resolution. They therefore offer an unprecedented opportunity to detect the signals of this multi-scale process directly from the kinematics of galaxies. The SAMI galaxy survey \citep{Bryant15} in particular already provides high-resolution kinematics for a substantial number of galaxies selected from the high completeness GAMA spectroscopic survey \citep{Driver11}, therefore allowing for the reconstruction of the underlying cosmic filaments in which these galaxies are embedded. The aim of this paper is to make use of kinematic parameters derived from SAMI IFS maps to explore the connection of the spin of galaxies to their large-scale anisotropic environment. 

For this purpose, we reconstruct the cosmic web across GAMA fields using a density field estimated through the mass-weighted tessellation of the distribution of galaxies. Real and synthetic samples are described in further details in Section~\ref{section:obs}, while the methods, especially methods of extraction of cosmic filaments, are described in Section~\ref{section:virtual}. Section~\ref{section:align} then shows that mass-dependent spin alignments predicted by simulations are also found in the SAMI survey. Section~\ref{section:sims} shows that results compare favorably with the Horizon-AGN hydrodynamical simulation \citep{Dubois14} and with the SHARK semi-analytical model of galaxy formation \citep{Lagos18}. Section~\ref{section:discuss} compares our results to similar recent observations. Section~\ref{section:conclusion} summarises our findings.

\section{Galaxy samples: selection, kinematics and classification}
\label{section:obs}

Let us describe the numerical and real datasets used for the analysis.

\subsection{Observed kinematic sample: the SAMI survey}
\label{section:sami1}

SAMI is a multi-object IFS mounted at the prime focus of the 3.9m Anglo Australian Telescope (AAT). It uses 13 state-of-the-art imaging fibre bundles, called hexabundles
\citep{BlandHawthorn11,Bryant11, Bryant12, Bryant14}. Each hexabundle is made out of 61 individual fibres with $1.6''$ angle and spans a 15'' diameter region on the sky. It has a maximal filling factor of $75\%$, and can be deployed over a $1^{o}$ diameter field of view. The 819 fibres, including 26 individual sky fibres, are fed into the AAOmega dual-beamed spectrograph \citep{Saunders04,Smith04, Sharp06}. The SAMI Galaxy Survey \citep{Croom12,Bryant15} sample consists of $\sim 3000$ galaxies across a broad range of galaxy stellar masses ($M_{*} = 10^{8}-10^{12}\,M_{\odot}$) and galaxy environment (field, groups, and clusters). The redshift range of the survey, $0.004 < z < 0.13$, corresponds to a spatial resolutions of 1.6 kpc per fibre at $z = 0.05$. Field and group targets were selected across the Galaxy and Mass Assembly (GAMA) G09, G12 and G15 regions \citep{Driver11} in four volume-limited galaxy samples derived from cuts in stellar mass. Cluster targets in SAMI are not used in our analysis.

For the SAMI Galaxy Survey, the 580V and 1000R grating are used in the blue ($3750-5750\, \AA$) and red ($6300-7400\, \AA$) arm of the spectrograph, respectively. This results in a resolution of $R_{\rm blue}\approx1810$ at $4800\,\AA$, and $R_{\rm red}\approx 4260$ at $6850\,\AA$ \citep{VdS17a}. In order to create data cubes with 0.5'' spaxel size, all observations are carried out using a six to seven position dither pattern \citep{Sharp15, Allen15}. Part of the reduced data-cubes and stellar kinematic data products in the GAMA fields are available on {\it https://datacentral.org.au}, through the first and second SAMI Galaxy Survey data release \citep{Green18, Scott18}. 

\subsubsection{Stellar kinematics}
\label{section:sami1}

Stellar masses are directly extracted from the GAMA catalog.
Precise kinematics measurements of the stellar intrinsic angular momentum, kinematic position angle and $v/\sigma$ within one effective radius (within elliptical aperture) are available for $\sim 60\%$ of SAMI galaxies across GAMA fields. Their computation was first described in \cite{VdS17b,VdS18}. We briefly summarize the method here.

Effective radii and ellipticities are derived using the Multi-Gaussian Expansion \citep{Emsellem94, Cappellari02} technique and the code from \cite{Scott13} on images from the $\rm GAMA-SDSS$, SDSS \citep{York00}, and VST \citep{Shanks13, Owers17}. $R_{\rm e}$ is defined as the semi-major axis effective radius, and the ellipticity of the galaxy is measured within one effective radius from the MGE best fit. 
Stellar kinematics are measured from the SAMI data by using the penalised pixel fitting code (pPXF)  \citep{Cappellari04}.

The fitting of galaxy cubes is performed with the SAMI stellar kinematic pipeline, which assumes a Gaussian line of sight velocity distribution (LOSVD), i.e. uses only the stellar velocity and velocity dispersion. The red spectra are first convolved to match the instrumental resolution in the blue. The blue and red spectra are then re-binned onto a logarithmic wavelength scale with constant velocity spacing ($57.9\,\rm km \rm s^{-1}$) with the pPXF package. We use annular binned spectra for deriving local optimal templates from the MILES stellar library \citep{SanchezBlazquez06}.

Once the optimal template is constructed for each bin, pPXF is run iteratively on each galaxy spaxel to measure the noise scaling from the fit residual (first pass), to mask emission lines and clip outliers (second pass), and finally to extract the velocity and velocity dispersion (third pass). For the latter, optimal templates from the annular bin in which the spaxel is located and from neighbouring annular bins are allowed. 

Eventually, the uncertainties on the LOSVD parameters are estimated from 150 simulated spectra. As detailed in \cite{VdS17b}, for the SAMI Galaxy Survey we impose the following quality criteria to the stellar kinematic data: signal-to-noise ($S/N>3$), $\rm obs> \rm FWHM_{\rm instr}/2= 35 \rm km \rm s^{-1}$ where the FWHM is the full-width at half-maximum, $V_{\rm error} < 30 \rm km \rm s^{-1}$, and $\rm error < \rm obs + 0.1 + 25 \rm km \rm s^{-1}$ (Q1 and Q2 from \cite{VdS17b}).

The kinematic position angle of the stellar rotation is measured from the two-dimensional stellar velocity maps on all spaxels with a $S/N>3$ and $V_{\rm ERR} < \rm 30 km.s^{-1}$ \citep{VdS17a}. The PAs are computed using the FIT KINEMATIC PA code, with the bi-antisymetrisation procedure described in Appendix C of \cite{Krajnovic06}. For the measurements, we assumed a centre of the map at (25.5,25.5) \citep{Scott18}. This $\chi^{2}$ minimization procedure allows to naturally produce individual $3\sigma$ errors on PAs as the $\chi^{2}=9$ contours. It is therefore straightforward to deduce the individual $1\sigma$ error on each PA. Such errors are typically slightly wider than bootstrap errors on the fit (resampling spaxels). PAs and errors on PAs display a gaussian distribution across the SAMI sample. Therefore, when resampling PAs is needed for the computation of significance contours, we draw every new individual PA from a gaussian distribution centred on the PA, with standard deviation matched to the individual $1\sigma$ error.

\subsubsection{Selection criteria and final sample}
\label{section:selec}

We start from the 2410 SAMI galaxies found across GAMA fields. We exclude all galaxies with $M_{*}<10^{9}\,M_{\odot}$ which have too few spaxels for a correct estimation to be made and do not have resolved counterparts in cosmological hydrodynamic simulations. 

Among the 1893 remaining galaxies, we select all those that have a measured PA, decreasing the sample to 1610 galaxies. This step typically excludes galaxies with too few spaxels and too low $S/N$ (generally low surface brightness galaxies). 

Note that PA errors are usually larger for very small (usually very low mass) galaxies and slowest rotators. The effect on very small galaxies is naturally corrected by the minimal mass threshold we impose in the next section. However, over-selecting fast rotators might not only enhance the low mass signal expected to be dominated by them, but more problematically erase the high mass signal suggested to be driven by slow-rotators \citep{Codis18}. We then further impose that the 1$\sigma$ uncertainty on PA should be $\rm \Delta_{\rm PA} < 25^{o}$, a very conservative fit which simply ensures convergence of the fits while preserving the morphological diversity in each mass bin. We are careful not to impose a stricter cut on PA precision to avoid excluding too many high mass slow-rotators which are crucial to the recovery of the expected alignment trend at high mass in simulations as the effect is maximal for them \citep{Codis18}. 

We also exclude merging systems which would lead to disturbances within 1 Re where PAs are calculated, but at this stage this removes only a handful of systems. This leads to a final number of 1418 usable stellar velocity and stellar velocity dispersion maps across GAMA fields.
The PA and $\Delta_{\rm PA}$ selection step has virtually no impact on most massive galaxies ($M_{*}>10^{11}\,M_{\odot}$) and very limited impact on galaxies with $10^{10}\,M_{\odot}<M_{*}<10^{11}\,M_{\odot}$, removing only $10\%$ to $15\%$ of such galaxies, spanning a wide range of effective radii $R_{\rm eff}$ and ellipticities. Only the minority of galaxies with $R_{\rm eff}>8 \, \rm kpc$, mostly found in the highest mass range, is entirely immune to the selection. The impact is the strongest on galaxies with $M_{*}<10^{9.5}\,M_{\odot}$, with 40 to 50$\%$ of them being removed by the cut, mostly small systems combining $R_{\rm eff}<3 \, \rm kpc$ and low surface brightness. The angular momentum of such systems is typically not resolved either in state-of-the-art simulations where alignments are predicted hence do not contribute to the signal.

Note that among the 1418 galaxies used in this study, those that have a measured $v/\sigma$ and ellipticity (1268) provide a good coverage of the $\lambda-\epsilon$ plane, with a distribution also consistent with that of the SAMI population with $M_{*}>10^{9}\,M_{\odot}$ and measured $v/\sigma$ when no PA cut is applied (1340). Hence the ranges of two main expected drivers of the correlations tested in this work, stellar mass and rotational support, are well preserved in our sample.

When specifically mentioned, we further impose that the effective radius and maximum measurable radius be bigger than the seeing HWHM to focus on most spatially resolved galaxies. The final number of galaxies in this case is 1278 across GAMA fields.
The corresponding stellar mass distributions can be found in Appendix.~\ref{section:mass-distr}.

\subsection{Observed spectroscopic sample: the GAMA survey}
\label{section:gama1}

To reconstruct the three-dimensional anisotropic large scale environment of our kinematic sample, in particular the cosmic nodes and filaments on mega-parsec scale, an accurate tracer of the underlying Mpc scale density field over a large volume (over a few 100 Mpc$^{3}$) is required. The DisPerSe software \citep{Sousbie11} can efficiently extract these cosmic features directly from a complete sample of galaxies with spectroscopic redshift measurements spanning a large volume without re-sampling. 

We use the GAMA spectroscopic survey sub-fields from which SAMI galaxies are selected (G09, G12 and G15). GAMA is a major campaign that combines a large spectroscopic survey of $300\, 000$ galaxies with most stellar masses comprised between $10^{8}$ and $10^{12}\, M_{\odot}$. This survey is carried out using the AAOmega multi-object spectrograph on the AAT. In the following study, we reconstruct the cosmic web from 2 different samples:
\begin{itemize}
\item The DR3 main survey sample \citep{Baldry18}. This is the latest data release of the GAMA survey and our reference sample. With about 100,000 robustly identified galaxies with secure redshifts and stellar mass information across the three sub-fields of interest, it is the most complete sample of the GAMA survey to date.
\item The DR3 main survey and fillers sample. With more than 119,000 objects usable in our reconstruction including a large number of faint low-mass fillers, this sample is the largest we use and potentially allows to probe filaments of smaller mass galaxies that would be otherwise considered as void galaxies. However, the completeness of the filler sample is much more limited than the main survey sample and the detection much less robust (only $12\%$ have secure redshifts with $nQ>2$, all of which are taken here but most of them have $nQ=3$). Hence this sample is the most likely to include spurious small scale filaments especially at higher redshift. It is nonetheless useful to check the robustness of the alignment trends of the lowest mass SAMI galaxies at low redshifts as it provides an increased resolution of cosmic filaments in regions where volume is lacking ($z<0.04$).
\end{itemize}
We find that results obtained with one or the other sample are similar both quantitatively and qualitatively. We therefore present only the DR3 main survey sample results hereafter.

\subsection{Simulated kinematic sample}

\subsubsection{Horizon-AGN}

The Horizon-AGN simulation is extensively described in \cite{Dubois14}. Here, we provide a very brief summary. This hydrodynamic cosmological simulation is run in a $L_{\rm box} = 100 h^{-1} \rm Mpc$ cube
with a $\Lambda$CDM cosmology with total matter density $\Omega_{\rm m}= 0.272$, dark energy density $\Omega_{\rm \Lambda}= 0.728$, amplitude of the matter power
spectrum $\sigma_{8} = 0.81$, baryon density $\Omega_{\rm b} = 0.045$, Hubble constant $H_{0} = 70.4 \,\rm km \rm s^{-1} Mpc^{-1}$, and $n_{\rm s}= 0.967 $ compatible with the WMAP-7 data \citep{Komatsu11}. The total volume contains
$1024^{3}$ dark matter (DM) particles, corresponding to a DM mass resolution of $M_{\rm DM,res} = 8 \times10^{7}\, M_{\odot}$, and initial gas resolution of $M_{\rm gas,res} = 1 \times10^{7} M_{\odot}$. It is run with the RAMSES code \citep{Teyssier02}, and the initially coarse $1024^{3}$ grid is adaptively refined down to $\Delta x = 1$  proper kpc, with refinement triggered if the number of DM particles in a cell becomes greater than 8, or if the total baryonic mass reaches eight times the initial baryonic mass resolution in a cell. Gas is heated by a uniform UV background after redshift $z_{\rm reion} = 10$ following \cite{HaardtMadau96} and can also cool down to $10^{4}\rm K$ through H and He collisions with a contribution from metals.

Star formation occurs where the gas number density is above $n_{\rm 0} = 0.1 \rm H cm^{-3}$ following a Schmidt law. Feedback from stellar winds, supernovae type Ia and type II are included with mass, energy and metal release, assuming a Salpeter initial mass function. The formation and coalescence of black holes (BHs), as well as consecutive AGN feedback in quasar and jet modes is also taken into account. More details can be found in \cite{Dubois14}.

\subsubsection{IFS mocks}

Galaxies were identified using the most massive sub-node method \citep{Tweed09} of the AdaptaHOP halo finder \citep{Aubert04} operating on the distribution of star
particles with the same parameters as in \cite{Dubois14}. Unless specified otherwise, only structures with a standard minimum of $N_{\rm min} = 50$ star particles are considered, which typically selects objects with stellar masses larger than $1.7 \times 10^{8} \,M_{\odot}$. Catalogs containing up to $150\, 000$ galaxies within the $10^{8.5}-10^{12.5} \, M_{\odot}$ stellar mass range are produced for each redshift output analyzed in this paper. Such computation of stellar masses is comparable to the way GAMA stellar masses are obtained in observations. The Horizon-AGN mass function at z=0 can be found in \cite{Canas18}.

For each galaxy with $M_{*}>10^{9.5}\,M_{\odot}$, we produce mock IFS kinematic maps projecting stars on a 1.5 kpc wide pixel square grid of adaptive size and fitting the mass-weighted velocity distribution in each pixel with a gaussian. In the present study, x-axis is taken to be the line-of-sight.
Kinematic parameters such as $\lambda$, $v/\sigma$ and higher order kinematic moments \citep{Ensellem07, Cappellari07,Krajnovic06} are later computed in elliptical aperture within a effective radius (circularised radius). Details of this procedure and morphological/kinematical comparison between mocks and SAMI data are studied in \cite{VdS18} and in further details in an upcoming paper (Welker et al., in prep.). 
In particular, the angular momentum (PA) of galaxies is computed directly from the kinematic maps within the half-light radius of galaxies following \cite{Krajnovic06} to optimize the comparison with SAMI results.

A major difficulty in comparing simulated galaxies in Horizon-AGN with SAMI galaxies is the resolution limit in the simulation. Galaxies in cosmological simulations are made of macro-particles of $10^{6}\,M_{\odot}$, accounting for populations of stars rather than a single star. Similarly the spatial resolution is limited to 1 kpc $h^{-1}$, which is the minimal size for gas cells, hence the minimal scale on which star formation and feedback processes can be computed. As a consequence, galaxies below $M_{*}<10^{8.5}\,M_{\odot}$ are not resolved while those with $M_{*}<10^{9.5}\,M_{\odot}$  are poorly resolved, with noisy, low-reliability shapes and angular momenta, especially in projection, when computed from IFS mocks directly comparable to the SAMI computation described in \cite{VdS18}. This implies that  an important fraction of low-mass galaxies in SAMI have no counterpart in Horizon-AGN.
However, it has better statistics than SAMI (around 150 000 galaxies per redshift snapshot) and a reasonably representative stellar mass function at all redshifts in the mass range considered.  In the following, selecting galaxies with $M_{*}>10^{9.5}\,M_{\odot}$ ensure they are resolved with at least 1000 star particles.

\section{Methods \& definitions}
\label{section:virtual}

Let us now review the methods used to produce virtual data sets, extract cosmic filaments and analyse galactic orientation.

\subsection{Extraction of the cosmic web}
\label{section:skeleton}

\subsubsection{GAMA filaments}

Starting from the full distribution of galaxies within all GAMA fields in which SAMI galaxies were selected, we reconstruct the cosmic web  as a network of cosmic filaments connecting nodes of the cosmic web where massive groups are typically found.

To perform this reconstruction, the 3D density field is reconstructed directly in redshift and real space from the distribution of galaxies using a mass-weighted Delaunay tessellation, which removes the need for direct smoothing that might impact the directions of filaments. The tessellated density field is then used as an input for the topology extractor DisPerSe \citep{Sousbie11}, which identifies the ridge lines of the density field to produce a contiguous network of segments that trace the spine of the cosmic web, i.e. the cosmic filaments. While no prior smoothing is applied, filaments are directly trimmed adaptively by DisPerSe according to a signal-to-noise criterium. Essentially, {\sc disperse} measures the robustness of a filament and trims the candidate catalogue in two ways: filament  persistence, the ratio of the value at the two critical points in a topologically significant pair of critical points (maximum-saddle, saddle-saddle or saddle-minimum); and local robustness, the density contrast between the critical points and skeleton segments with respect to background. Removing low-persistence pairs is a multi-scale non-local method to filter noise/low significance filaments. When applied to point-like distributions of haloes or galaxies, a persistence threshold translates easily into a minimal signal-to-noise ratio, expressed as a number of standard deviations $\sigma$. This focus on topology rather than geometry to identify and estimate the reliability of filaments is one of the advantages of Disperse, allowing to recover a true multiscale network of filaments robust to re-sampling. This algorithm has not only been used to analyse simulations \cite[e.g.][]{Dubois14,Welker14,Welker18,Codis18,Kraljic18b} but has also been successfully applied to real spectroscopic (VIPERS, GAMA) and photometric (COSMOS) surveys on wider redshift ranges \cite[e.g.][]{Malavasi17,Laigle18,Kraljic18}. 

For the GAMA DR3 main survey sample, we produce the $2\sigma$ and $3\sigma$ networks of filaments. By construction, the $2\sigma$ skeleton, displayed in Appendix~\ref{section:refine}, contains fainter, less robust filaments. This allows to probe fainter, on average smaller scales of the cosmic web with the caveat that noise is significantly increased. It allows to test the expectation that the transition mass of spin alignments should vary from faint tendrils to highly collapsed, larger scale cosmic filaments  \citep{Codis15,Cautun15} (see Appendix.~\ref{section:refine}).

\begin{figure*}
\center \includegraphics[width=2\columnwidth]{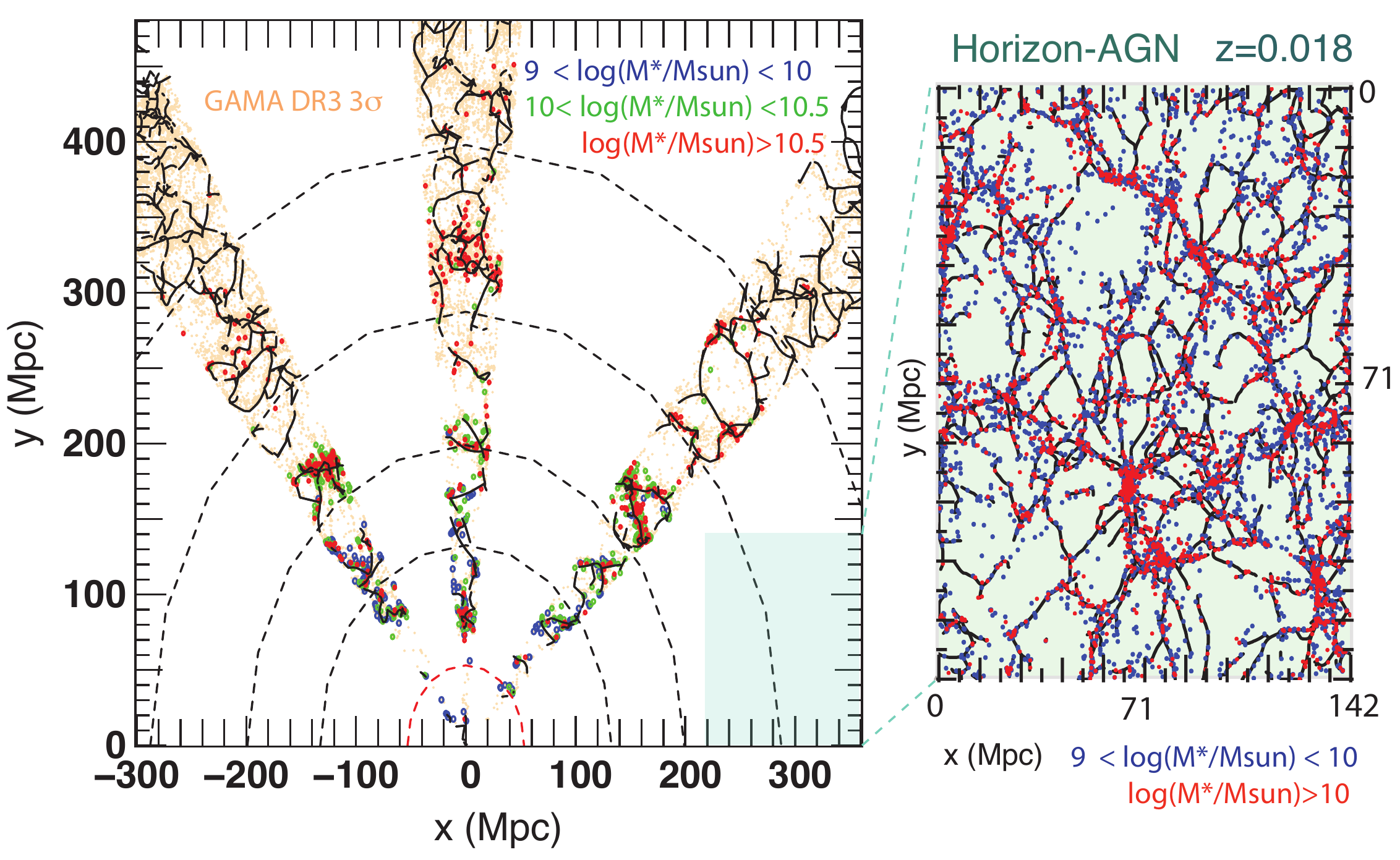}
  \caption{Projected reconstructed network of cosmic filaments (deprived of filaments aligned with the line-of-sight i.e. with $|\cos \alpha_{\rm los}|>0.95$) across the three GAMA fields that host SAMI galaxies (solid black lines). SAMI galaxies with $10^{9}\, M_{\odot}<M_{*}<10^{10}\, M_{\odot}$, $10^{10}\, M_{\odot}<M_{*}<10^{10.5}\, M_{\odot}$ and $M_{*}>10^{10.5}\, M_{\odot}$ are indicated as red, blue and green circles respectively. Light pink dots indicate the initial GAMA population from which the filaments are extracted. Dashed hemicircles indicate the redshift tiers of the SAMI survey. The right inset shows the galaxy population and gas filaments extracted in a 25 Mpc thick slice from the Horizon-AGN simulation for comparison. }
\label{fig:cw}
\end{figure*}

The projected $3\,\sigma$ skeleton for the DR3 sample, i.e. the network of filaments, across GAMA fields G09, G12 and G15 is presented in black in Fig~.\ref{fig:cw}. Note that while we reconstruct the skeleton across the entire GAMA field (pink dots), SAMI galaxies are overlaid on top of the filaments as larger red, blue and green circles. Blue circles indicate SAMI galaxies with $10^{9}\, M_{\odot}<M_{*}<10^{10}\, M_{\odot}$, green circles those with $10^{10}\, M_{\odot}<M_{*}<10^{10.5}\, M_{\odot}$ and  blue circles those with $M_{*}<10^{10.5}\, M_{\odot}$. Dashed hemicircles indicate the redshift selection tiers of the SAMI survey ($z=0.03, 0.045, 0.06, 0.095$). 
The effect of the mass-redshift selection tiers on the environment sampling is clearly visible on this picture: beyond the two lowest redshift tiers there is virtually no galaxy with $M_{*}<10^{10}\, M_{\odot}$, while in the third tier, the SAMI selection contains galaxies in the vicinity of denser nodes. To limit biases, we checked that despite a high level of noise due to low numbers, our results remained consistent in each redshift tier. 

\begin{figure}
\center \includegraphics[width=0.9\columnwidth]{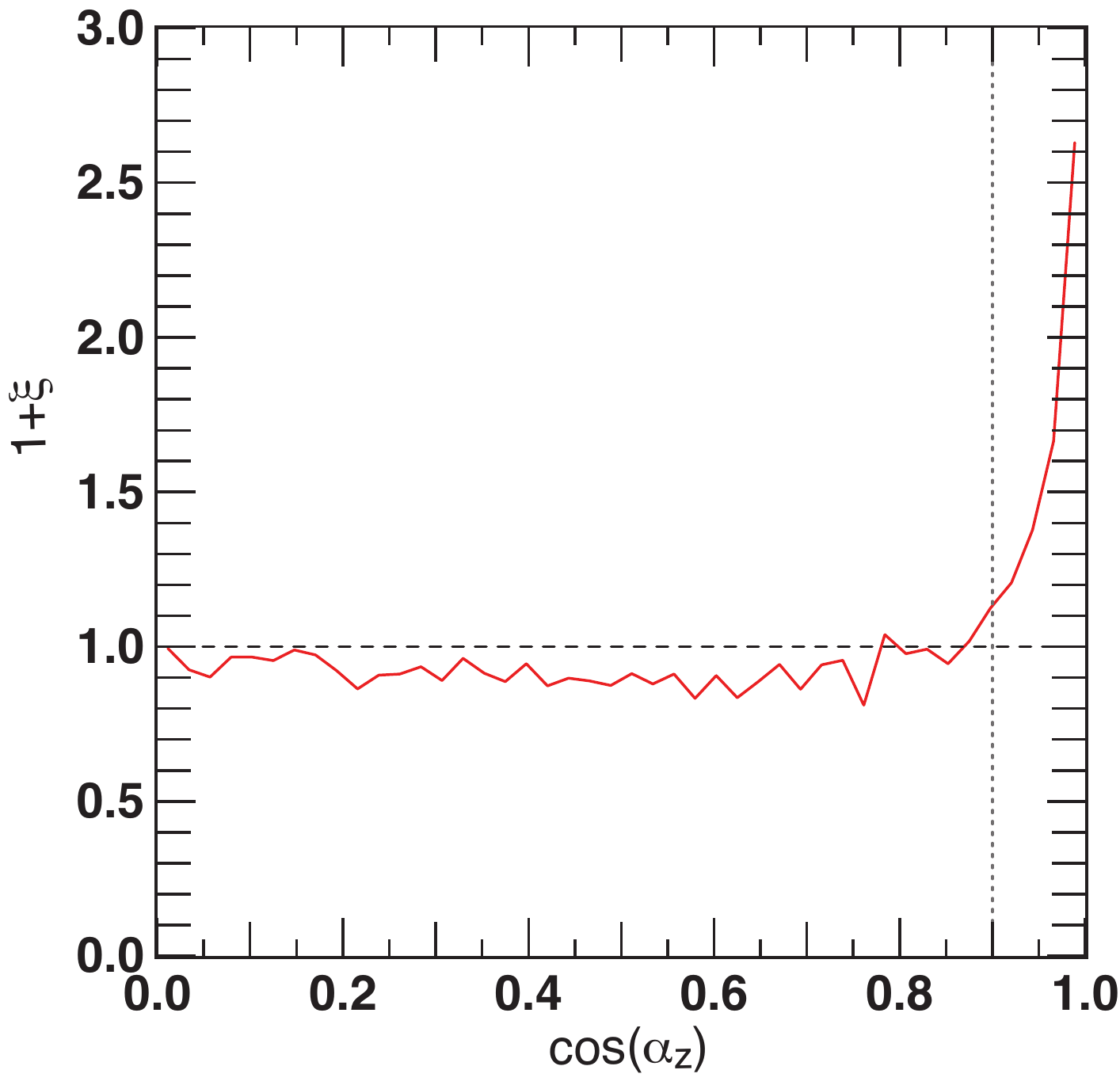}
  \caption{PDF of $\cos(\alpha_{\rm los})$, with $\alpha_{\rm los}$ the angle between a given filament segment and the line of sight in the GAMA fields prior to any correction. An excess of aligned segments is seen past $\alpha_{\rm los}\approx 0.9$. This amounts to around $5\%$ of the filaments. The threshold in black shows the cut used to identified contaminated galaxies.}
\label{fig:fog-pdf}
\end{figure}

In spectroscopic surveys, a major source of concern is redshift space distortion effects, in particular ``finger-of-god" effects. This can artificially stretch groups along the line of sight, resulting in the creation of spurious filaments in that direction. This is easily seen on Fig.~\ref{fig:fog-pdf} which shows the PDF $1+\xi$ of the absolute value of the cosine of the angle $\alpha_{\rm los}$ between the line of sight and filament segments in redshift space (green curve). The expected uniform distribution is shown as a dashed line for comparison with $\xi$, the excess probability (from uniform). While the PDF is near-uniform across a large range of angles, an excess of segments along the line of sight is clearly detectable as an excess probability in $|\cos \alpha_{\rm los}|>0.9$, especially pronounced for $|\cos \alpha_{\rm los}|>0.95$ with $\xi>2.5$. This corresponds to $6\%$ segments with $0.9<|\cos \alpha_{\rm los}|<0.95$ and $10\%$ segments with $|\cos \alpha_{\rm los}|>0.95$ instead of the $5\%$ expected in each bin from the random distribution. Note however that a small degree of anisotropy is also expected due to the elongated geometry of the survey.

Several technics can be used to correct for such distortion effects. In particular, \cite{Kraljic18} chose to identify richest groups in the GAMA field using an anisotropic Friend-of-friend algorithm, then shift the positions of hosted galaxies so as to reset the LOS dispersion equal to the transverse dispersion prior to extracting the skeleton. We choose not to use this technic for several reasons:
\begin{itemize}
\item This can only be efficiently applied to rich enough groups ($\approx 10$ members), where the finger-of-god effect is maximal. Our focus is on the very low-redshift part of the GAMA fields ($0.01<z<0.1$) where such groups are under-represented due to limited volume. This issue is thus limited but harder to correct for in our regime.
\item While it is true that groups' tranverse and LOS dispersions should be statistically equal, this does not apply to individual groups which generally display a degree of stretching and infall along their cosmic filament (resulting in a confinement of the orbits), which can be arbitrarily aligned of misaligned with the line of sight. Our very limited statistics require to take this effect into account.
\item This technique does not correct for boundary effects that also typically produce spurious filaments along the line of sight.
\end{itemize}

To account for finger-of-god effects, we first flag the filaments with  $|\cos \alpha_{\rm los}|>0.9$ and $|\cos \alpha_{\rm los}|>0.95$. Then we apply three correction methods:
 
\begin{itemize}
\item {\bf Method 0}:  We do not apply any correction. All galaxies are taken into account but this tend to underestimate the distance of massive group galaxies to filament.
\item {\bf Method 1}:  For each SAMI galaxy assigned to a segment with $|\cos \alpha_{\rm los}|>0.9$, we identify the next contiguous filament segment with $|\cos \alpha_{\rm los}|<0.9$, the galaxy is re-assigned to it. All galaxies are taken into account but this tend to overestimate the distance of massive group galaxies to their nearest filament.
\item {\bf Method 2}:  We disregard all SAMI galaxies assigned to a segment with $|\cos \alpha_{\rm los}|>0.95$.
\end{itemize}

We find that our results are qualitatively independent of the method applied, with only minor differences in the amplitude of the signal and the noise. In particular, as could be expected, the sharp cut performed in Method 2 leads to an increase of the signal but to a even stronger increase of the level of noise due to the sharp cut on statistics. This limited effect of finger-of-god distortion is expected as, while we assign SAMI galaxies to their GAMA filaments in real space, angles are computed in projection, hence any component along the line-of-sight is not taken into account. The major parameter that could be impacted is the distance to the nearest filament for galaxies in groups (as this filament might spuriously extend along the line of sight), but the limited span of this effect on our low-mass groups does not modify the signal on the scales on which we resolve the cosmic web.  
Unless specified otherwise, results are presented using Method 1 in what follows.

\begin{figure}
\center \includegraphics[width=0.9\columnwidth]{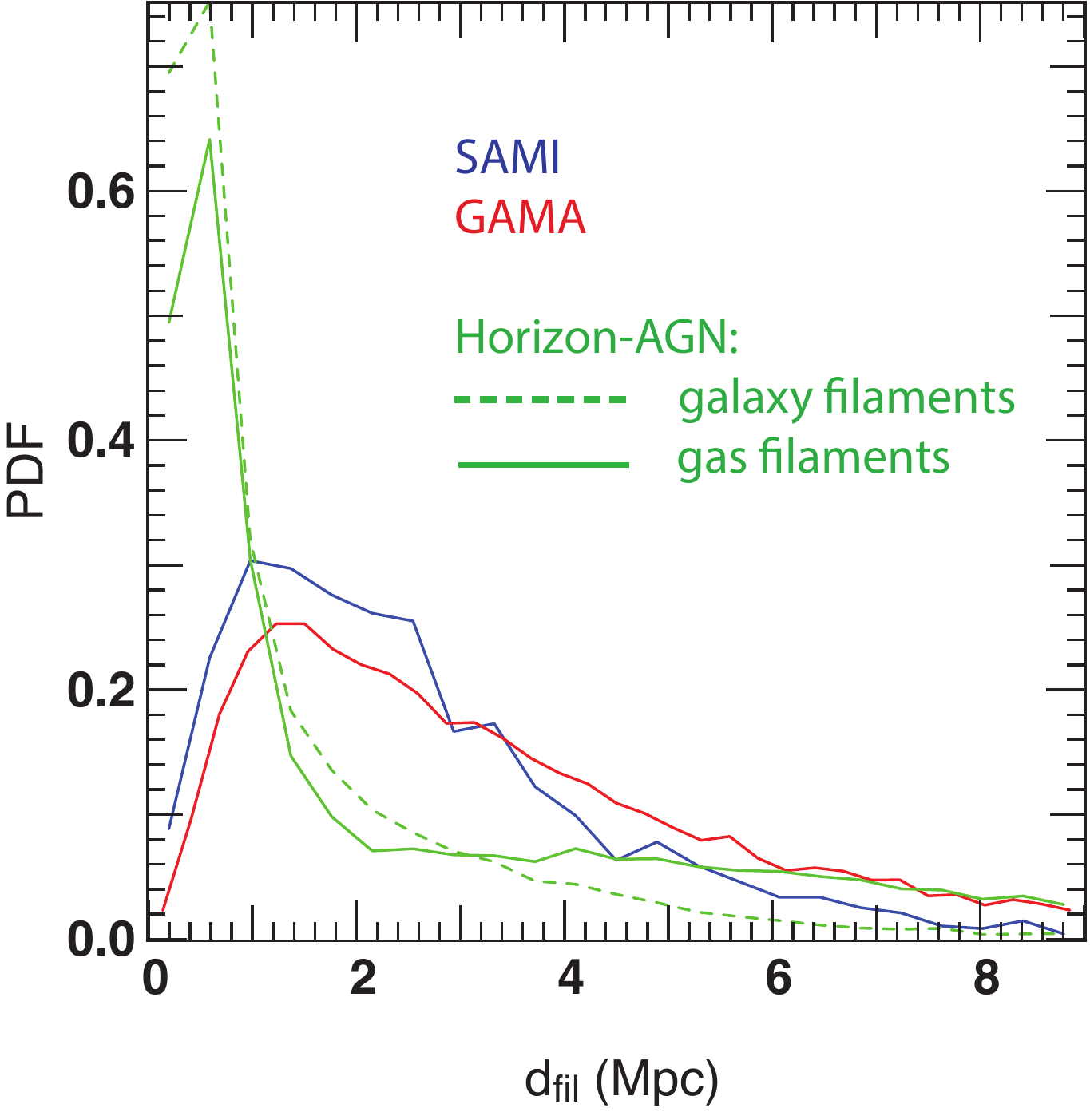}
  \caption{PDF of the distance to the nearest filaments for observed and simulated galaxies. The PDF for all GAMA galaxies in the redshift range considered ($0<z<0.1$) appears in red, that for all SAMI galaxies used in this study in blue, and that for Horizon-AGN galaxies in green. While similar in terms of median and average, effects of completeness and redshift distortions increase the dispersion around filaments for observed galaxies compared to simulated ones.}
\label{fig:dist}
\end{figure}

Fig.~\ref{fig:dist} displays the PDF of the un-smoothed distance to the nearest filaments for observed (red and blue) and simulated galaxies (in green, this aspect is developed in the next section). The red histogram displays the PDF for all GAMA galaxies with $M_{*}>10^{9}\, M_{\odot}$ in the redshift range considered ($0<z<0.1$), the blue histogram the PDF for the full SAMI sample selected in this study (also $M_{*}>10^{9}\, M_{\odot}$). The presence of a peak offset from the filament spine is due to the discreteness of the sample used to extract filaments and the position of the peak is therefore a good indicator of the uncertainties on filament positions. Indeed, since filaments are extracted from a point-like distribution of galaxies tessellated to estimate a local density  by construction they run through the point-distribution, which means no galaxy can be found at the spine of such filaments unless it is at a node of the network. Due to this fitting procedure, galaxies naturally distribute around it and the peak offset therefore traces the uncertainty on the position of the filament due to the mean intergalactic distance.

One can see that the SAMI sample displays a reasonably representative distribution of galaxies around the spine of the cosmic web as compared to GAMA. The deficiency of ``void" galaxies with $d_{\rm fil}>4\, \rm Mpc$ (with $d_{\rm fil}$ the distance of a galaxy to its nearest filament) and the stronger peak around 1 Mpc are expected as a characteristic of the over-representation of high-mass galaxies in the sample. The shape and standard deviation of the distribution are nonetheless overall preserved. Despite its limited number, the SAMI sample displays a representative distribution around cosmic structures, which is an important feature of this sample for the kind of analysis presented here. Indeed, in theory and simulations the evolution of the spin orientation in the cosmic web is driven by to the progressive migration of galaxies closer to the spine of the filament as they grow in mass. Failure to properly reproduce this basic feature would therefore greatly limit our ability to detect a signal. 
Although patchy across the GAMA fields, it is important to stress that the SAMI selection was designed so as to probe a variety of cosmic structures (field, pairs, groups).

\begin{figure}
\center \includegraphics[width=0.9\columnwidth]{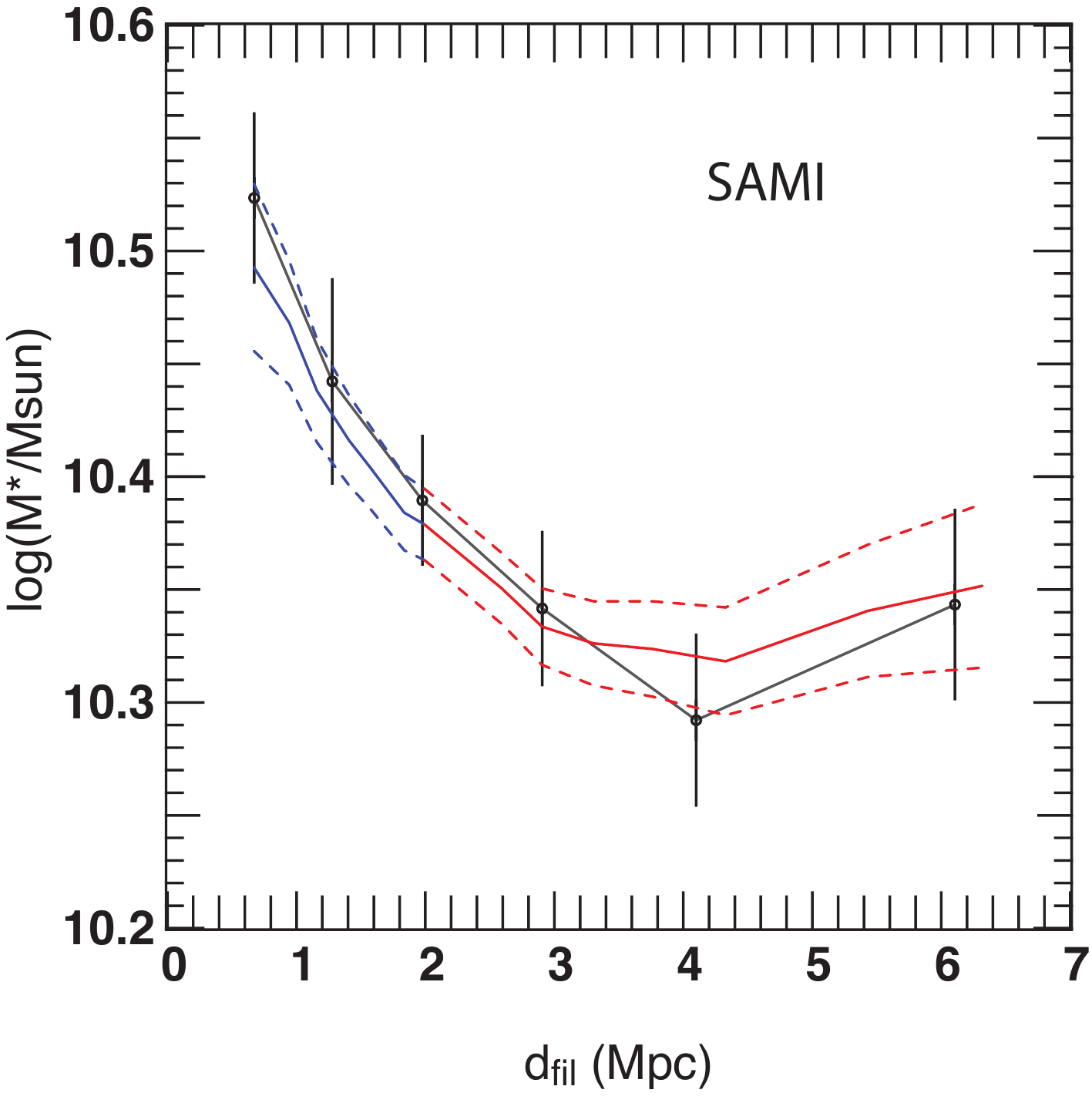}
  \caption{ Average stellar mass as a function of distance to filament for SAMI galaxies for the full sample using the two sets of overlapping bins (blue and red lines) described in the text and using standard contiguous bins (black line). The steady decrease of average stellar mass with distance to filament is recovered across the full sample. Error bars show 1-$\sigma$ errors on the mean.}
\label{fig:mdfil}
\end{figure}

Moreover, Fig.~\ref{fig:mdfil}, left panel shows that the average evolution of stellar mass with distance to filament for the full SAMI sample is representative of what is expected from theory and simulations (see Appendix.~\ref{section:SAMI2}.). The black line displays the results using standard contiguous bins. To overcome low number statistics, we also define irregular, overlapping distance bins but we ensure that the median distance (in a bin) increases smoothly and steadily across consecutive bins. To do so we consider:
\begin{itemize}
\item ``Blue" bins: each bin contains all galaxies with $d_{\rm fil}$ between 0 Mpc and  $d_{\rm i}$ with $d_{\rm i}$ taking values 1, 1.5, 2, 3, 4, 5, 6  and 7 Mpc.
\item ``Red" bins: each bin contains all galaxies with $d_{\rm fil}$ between $d_{\rm i}$ and 7 Mpc with $d_{\rm i}$ taking values 0, 1, 1.5, 2, 3, 4, 5, 6 and 7 Mpc.
\end{itemize}
Then for each bin, we computed the average distance to filament for the galaxies it contains and plotted the dependence of stellar mass with $d_{\rm fil}$. We checked in the Horizon-AGN simulation that this binning method preserves the trend obtained using regular distance bins when larger statistical samples are available (see Appendix.~\ref{section:SAMI2}). This method tends to decrease error bars in a given bin, at the expense of independence between bins. Effectively, it smoothes out the curve more local environmental effects (integrative-like smoothing). The case of strong steady signals as that in Fig.~\ref{fig:mdfil} shows that the match with the signal obtained with regular bins is excellent. In the following study, when specifically mentioned, we will make use of similar binning methods to identify robust features such as transition masses in otherwise noisy trends.

The steady decrease of average stellar mass with distance to filament is recovered for the full sample. More massive galaxies tend to cluster closer to the spine of the filament than their less massive, younger counterparts. This result is consistent with studies in the full GAMA survey \citep{Kraljic18} and in high redshift photometric surveys \citep{Malavasi17,Laigle18} and favours a scenario of progressive migration of galaxies along the cosmic web as they accrete gas through diffuse accretion and mergers and therefore grow in mass. 

We further checked that despite variations on the average mass and environment, the observed trend is also recovered in every redshift tier individually and that a similar feature is recovered when analyzing distance to nodes of the cosmic web. Overall, the SAMI sample is very representative of both its parent spectroscopic survey and numerical predictions in terms of number count and mass distributions around cosmic filaments and is therefore particularly well suited for the analysis of the cosmic connection between filaments and galactic spins.

\subsubsection{Horizon-AGN simulation filaments}

To allow for a direct comparison with simulations, we also extract the cosmic web in the Horizon-AGN simulation using the same software. We do it in two different ways:
\begin{itemize}
\item {\it Galaxy filaments}: Similarly to what was done in GAMA, we use the distribution of galaxies identified in Horizon-AGN and we extract the $3\sigma$ cosmic web. We restrict the sample to galaxies with $M_{*}>10^{9}\,M_{\odot}$ to limit the impact of the under resolved low mass galaxies which show strong departure from the expected $z=0$ mass function in Horizon-AGN. Note however that we use the full cubic box and not the Horizon-AGN lightcone \citep{Laigle19} which has too limited volume at low redshift, hence the geometry of the survey is not further fitted to GAMA. Resampling Horizon-AGN to fit the GAMA mass function did not lead to any significant differences with the $3\sigma$ cut.
\item {\it Gas filaments:} We extract the filaments directly from a gas density cube of the full Horizon-AGN volume, with a cell size of $200 \rm kpc\,\rm h^{-1}$. Using the gas filaments provides an additional check that filaments extracted in GAMA -- and consecutive properties of galaxies around them -- are consistent with a direct detection of simulated density filaments. We stress that in the simulated volume we have in theory access to smaller-scale, lower density gas filaments. As this could limit the comparison, we adapt the level of persistence  to reproduce only the class of filaments detectable in GAMA, that is filaments with an extent of order 10-15 Mpc on average. This is done by checking that the average and median inter-critical point distances (distances between nodes and filament type saddle points) are of the same order of magnitude for the GAMA and the Horizon-AGN gas skeleton. In Horizon-AGN, gas filaments have a median inter-critical points distance of 3.2 Mpc. while it is 5.2 Mpc for the GAMA $3\sigma$ skeleton and 4.5 Mpc for the $2\sigma$ skeleton. Variations are expected mostly due to the different geometries of the GAMA survey and the Horizon-AGN simulation.
\end{itemize}

\begin{figure}
\center \includegraphics[width=0.85\columnwidth]{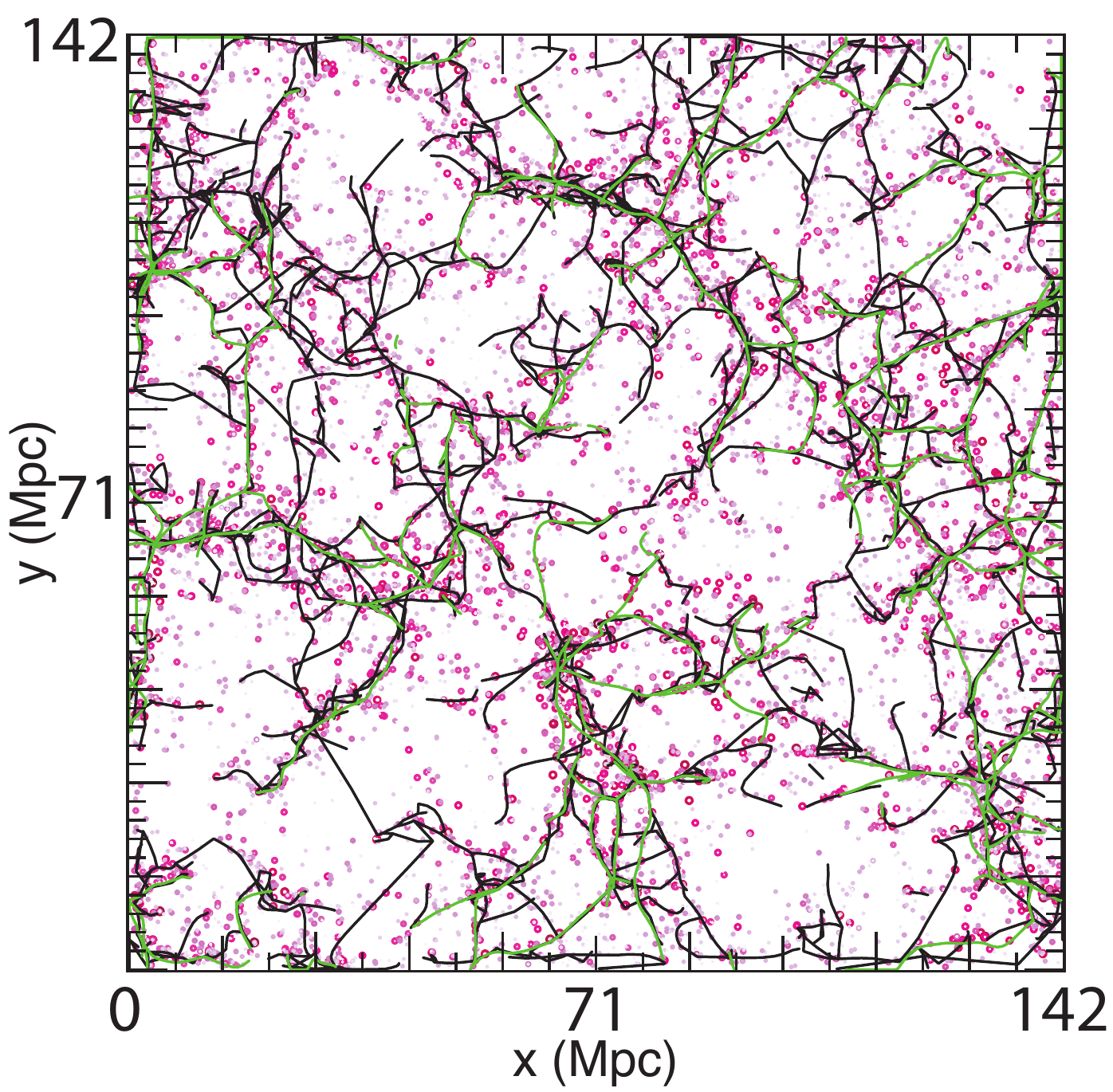}
  \caption{ Map of a 25 Mpc thick projection of galaxies (dots in shades of pink) and cosmic filaments extracted from galaxies (black lines) and from gas density (green lines) in Horizon-AGN. Darker shades of pink indicate higher stellar masses. Using the galaxies allows to recover filaments that are well constrasted in terms of galaxy distribution and seem relevant in the build-up of galaxies but have become too faint in terms of gas density. }
\label{fig:maphagn}
\end{figure}

Fig.~\ref{fig:maphagn} displays a 25 Mpc thick projection of galaxies (dots in shades of pink) and cosmic filaments extracted from galaxies (black lines) and from gas density (green lines) in Horizon-AGN. Darker shades of pink indicate higher stellar masses.
The difference between the two is striking. While most contrasted filaments are consistent in both cases, using the galaxies allows to extract thinner tendrils of galaxies that are below the persistence level used for gas filaments (given the smoothing imposed to the gas cube). This highlights the difference between a multiscale galaxy-based network and a network of similar average filament length but obtained from a smoothed field. While decreasing the persistence level in the gas filaments extraction allows to recover most of these tendrils, it also reveal many more gas tendrils. To summarize, using the galaxies allows to recover filaments that are well contrasted in terms of galaxy distribution and seem relevant in the build-up of galaxies but have become too faint in terms of gas density. Note however that the use of galaxies, even with a $3 \sigma$ persistence threshold also produce some spurious filaments. This is more pronounced for very massive galaxies in particular, which are therefore more likely to be classified as node galaxies, limiting our ability to assign them to one single relevant filament. The use of these two distinct networks of filaments for the comparison allows to check the robustness of the signal with scales and tracers of the cosmic web.

We produce a network of filaments for the five simulation snapshots with redshifts within the SAMI range ($z=0.018,0.036,0.055,0.075,0.095$). However, the network of filaments and the population of galaxies show very little evolution over this redshift range. 
We further check the compatibility of such filaments by ensuring that galaxies distribute similarly around them in both cases. As an example, Fig.~\ref{fig:dist} displays the PDF of the distance to the nearest filaments for observed galaxies and simulated galaxies (in green) for the gas filaments (solid line) and galaxy filaments (dashed line). One can see that observed and simulated populations display broadly similar trends around filaments: both the GAMA and Horizon-AGN PDFs peak between 500 kpc and 1.2 Mpc then decrease and display persistent tails away from filaments ($>4$ Mpc). Skeleton persistence for gas filaments is chosen so that they have medians comprised within 1 Mpc:  $d_{\rm med}=2.7\, \rm Mpc$ in GAMA, $d_{\rm med}=1.74\, \rm Mpc$ for the galaxy filaments in Horizon-AGN and $d_{\rm med}=1.78\, \rm Mpc$ for the gas filaments Horizon-AGN. Averages for GAMA and Horizon-AGN are also comprised within 1 Mpc  (3.2 Mpc to 3.8 Mpc respectively). It should be noted that differences in mass functions used (full Horizon-AGN or mocking the GAMA mass function) did impact strongly the distribution of simulated galaxies around filaments. The distribution for the GAMA mock was only found to be slightly more peaked towards the spine of filaments due to the over-representation of massive galaxies compared to the full Horizon-AGN sample.

The main differences between real and mock distributions highlighted in Fig.~\ref{fig:dist} include the stronger peak of the Horizon-AGN samples (and consecutive deficiency between 1 and 4 Mpc) compared to observations. This is expected as simulated galaxies do not suffer from varying completeness and uncertainties in mass and redshift, hence are more tightly packed around filaments. On the observed filaments, the spread in redshift measurements translates into an increased spread in distances to filaments along the line of sight, an effect visible on Fig.~\ref{fig:cw}. The long-distance tail of the synthetic distribution ($d_{\rm fil}>5\, \rm Mpc$) is also more pronounced for filaments extracted directly from the gas density field, i.e. independently from the galaxy distribution, and not inferred from the galaxy point distribution itself as is done in observations. 

\section{Spin-filament alignments in SAMI.}
\label{section:align}

\subsection{Alignments away from clusters.}
\label{section:field}

In order to estimate the degree of alignment of galaxies kinematic axis with their nearby filament, we first assign each SAMI galaxy to its nearest filament segment in the three dimensional filamentary network reconstructed from overlapping GAMA field (using the smallest 3D euclidian distance). We then compute the angle $\theta_{\rm kin}^{\rm 2D}$ between the kinematic spin axis position angle of each galaxy and the filament segment projected on the celestial sphere. Throughout this paper, we consider only these ``apparent" angles between a galaxy's spin and the projected direction of its associated filament. Apparent angles, projected on the celestial sphere, are not to be confused with true angles in 3D.

\begin{figure}
\center \includegraphics[width=0.9\columnwidth]{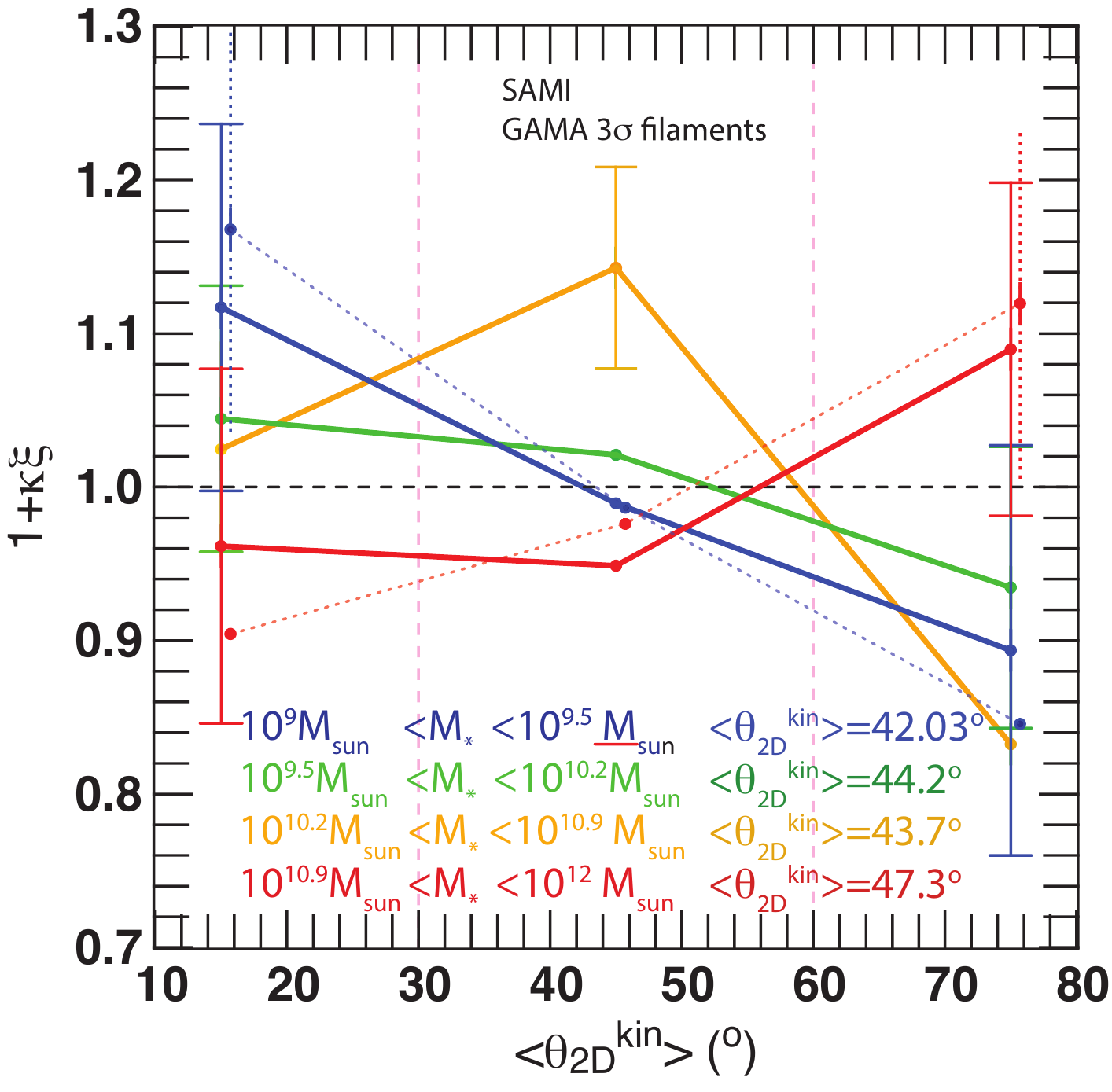}
  \caption{Renormalized PDF of $\theta_{\rm 2D}$, the angle between the kinematic axis of galaxies and their nearest filament, for galaxies with $10^{9}\, M_{\odot}<M_{*}< 10^{9.5}\, M_{\odot}$ (in blue), with $10^{9.5}\, M_{\odot}<M_{*}< 10^{10.2}\, M_{\odot}$ (in green), with $10^{10.2}\, M_{\odot}<M_{*}< 10^{10.9}\, M_{\odot}$ (in orange) and with $10^{10.9}\, M_{\odot}<M_{*}< 10^{12}\, M_{\odot}$  (in red). Dotted curves show the the PDF in the two extreme bins using Method 2 for redshift corrections. The dashed black line indicate expectation for a uniform distribution.Vertical pink dashed lines indicate the boundaries of the 3 angle bins used to produce this plot.}
\label{fig:align1}
\end{figure}

Fig.~\ref{fig:align1} shows the renormalized PDF  $1+\kappa\xi$ -- where $\xi$ is the excess probability and $\kappa$ a renormalisation factor  -- of $ \theta_{\rm kin}^{\rm 2D}$ for the SAMI sub-sample with $M_{*}>10^{10.9}\,M_{\odot}$ (in red), with $10^{10.2}\, M_{\odot} <M_{*}< 10^{10.9}\, M_{\odot}$ (in orange), with $10^{9.5}\,M_{\odot}<M_{*}< 10^{10.2}\, M_{\odot}$ (in green) and for the sub-sample with $10^{9}\, M_{\odot}<M_{*}< 10^{9.5}\, M_{\odot}$ (in blue). The expected signal for uniformly distributed angles is shown as a horizontal dashed black line. The renormalisation factor $\kappa=90$ is set so as to have the uniform distribution set to 1 (i.e. $\kappa \rm PDF= 1+\kappa\xi$), making it more comparable to similar analysis in 3D (therefore using the cosine rather than the angle) in simulations \citep{Dubois14,Welker14,Codis12,Codis18}. Angle bins are indicated in light pink vertical dashed lines. In addition, dotted red and blue curves show the same PDF in the two extreme bins using Method 2 rather than 1 for redshift corrections. They are slightly shifted horizontally to allow for better visibility. For the same reasons, error bars are omitted in some angle bins as they show little variation across a given PDF. Note that similar results are obtained using Method 0.

Focusing on the blue curve (low-mass galaxies), we observe an excess probability $\xi=0.12\pm0.12$ for $\theta_{\rm kin}^{\rm 2D}<30^{\rm o}$, hence an excess of alignment between such galaxies and their nearby filament. Similarly, a deficit of more perpendicular orientation is found in this mass bin with $\xi=-0.11\pm0.13$ for $\theta_{\rm kin}^{\rm 2D} > 60^{\rm o}$. Focusing on the red curve (high-mass galaxies), the opposite trend is found with an excess of galaxies displaying a kinematic spin axis orthogonal with $\xi=0.1\pm0.12$ for $\theta_{\rm kin}^{\rm 2D} > 60^{\rm o}$ to their nearby filament and a deficit of aligned orientations ( $\xi=-0.04\pm0.11$ for $\theta_{\rm kin}^{\rm 2D}<30^{\rm o}$). Note that in each case, what constitutes a clear interpretable signal is a roughly monotonous evolution combining an probability excess at one extreme of the cosine range and an excess of opposite sign at the other extreme. Intermediate results are found for the intermediate mass bins, highlighting the progressive transition of the spin orientation with increasing mass.  Similar results are found using the $2\sigma$ filaments, although the maximal signal is obtained for a reduced transition mass: galaxies as low as as $M_{*}>10^{10.2}\,M_{\odot}$ already display a preferentially orthogonal orientation of the spin. This is consistent with the idea that the transition mass is measured with respect to the mass of non-linearity hence depends on the scale of filaments, with thinner fainter filaments leading to lower transition masses. $2\sigma$ filaments results are presented in Appendix.~\ref{section:refine}.

This is consistent with the scenario suggested in \cite{Codis12} and later developed in \cite{Dubois14}, \cite{Welker14} and \cite{Laigle15}. Using N-body and hydrodynamic cosmological simulations, these studies show that low-mass galaxies are formed in the mid-filament region, offset from its spine, in quadrants of coherent vorticity (whirling flows) aligned with the filament direction as a result of anisotropic tides from the proto-filament embedded in its proto-wall at early cosmic times. In such regions, galaxies accrete coherent, high angular momentum gas and build up their spin parallel to the filament. As they grow in mass, they migrate towards the spine of the filament, then along the filament towards nodes of the cosmic web. During this phase, they accrete matter from regions that overlap vorticity quadrants of opposite polarity (hence destroying the coherence of the accretion). Galaxies also undergo mergers along the filament, resulting in a transfer of pair orbital momentum to the intrinsic angular momentum of the remnant. This tends to flip galactics spins orthogonal to their nearby filament.

Using more than $150,000$ galaxies, \cite{Dubois14} and \cite{Codis18} predict that such a signal is robust and detectable using various tracers of galaxy evolution, yet very weak in the redshift ranges  $z=1.2-1.8$ and $z=1-0$, with typical excess probabilities $\xi<0.05$. Most of these predictions however include galaxies in massive groups and clusters, where processes such as dynamical friction and local torques on satellites decrease the signal significantly, as highlighted in \cite{Dubois14}. 

\subsection{Amplitude of the excess alignments}
\label{section:ampl}

To test our signal against spurious noise, we use two distinct methods:
\begin{itemize}
\item we re-assign galaxies to random filament segments in our network and we compute the same PDF for 100 000 such samples.
\item we randomly flip galactic position angles by $45^{\rm 0}$ or $-45^{\rm 0}$ while keeping them assigned to their actual nearest filament and repeat this re-sampling procedure 100 000 times.
\end{itemize}
We then segment the full SAMI sample into a low-mass and high mass subsample using a mass threshold variable $M_{\rm thresh}$.

Note that in the following, we also fully take into account the individual errors on PAs. For each, realization, each galaxy PA is re-drawn from a gaussian distribution centred on the original PA, with a standard deviation equal to the individual $1\sigma$ error on the PA. This is however a very sub-dominant contribution to the significance contours, largely dominated by the low number statistics.

\begin{figure*}
\center \includegraphics[width=1.9\columnwidth]{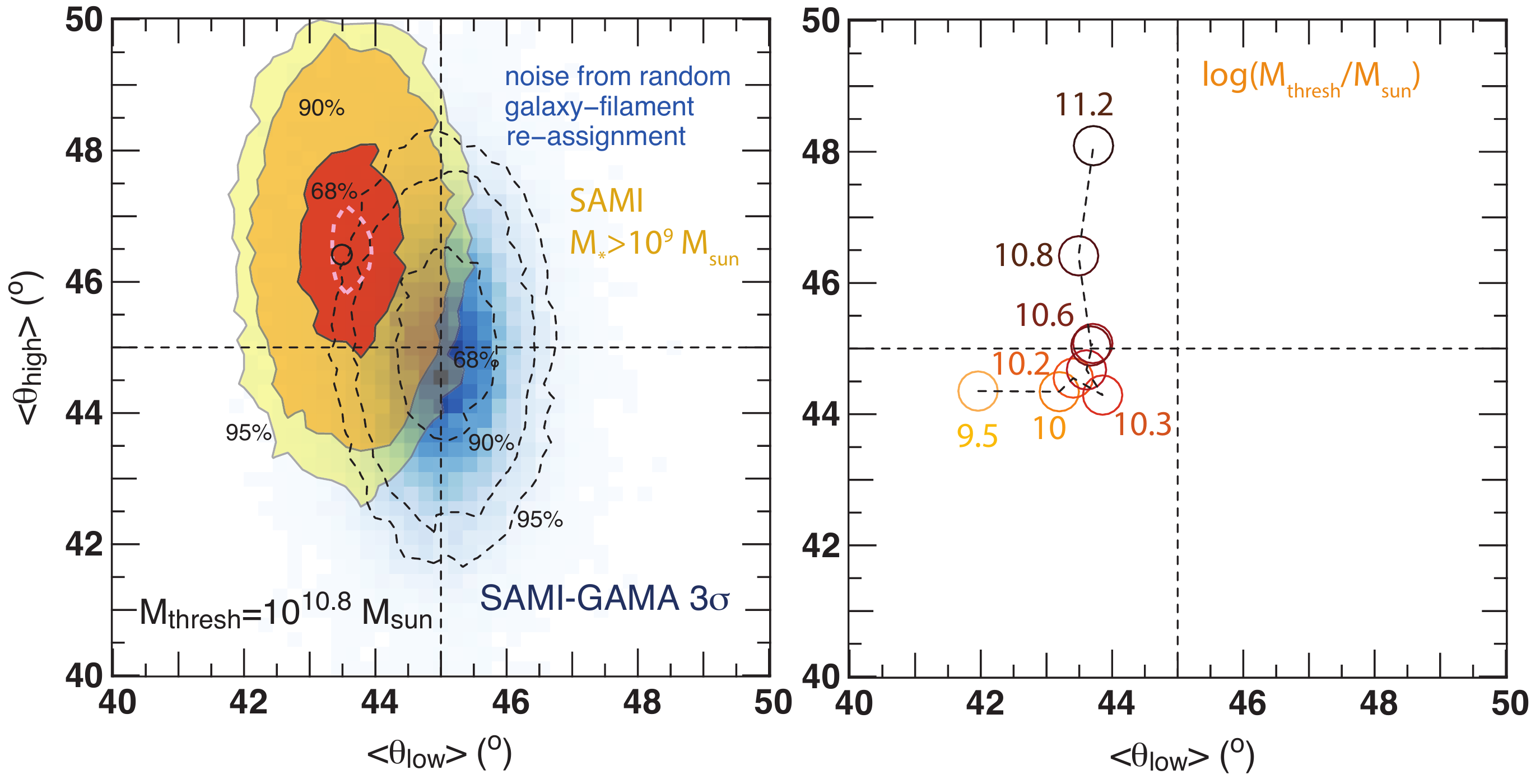}
  \caption{{\it Left panel}: Average spin-filament angle for galaxies with $M_{*}< 10^{10.8}\, M_{\odot}$ (``low") versus average angle for galaxies with $M_{*}> 10^{10.8}\, M_{\odot}$ (``high") using redshift distortion correction method 1. Shades of blue and dashed contours indicate the distribution of values for the expected level of noise, obtained from random re-pairing of galaxies and filaments. Straight black dashed lines show the expectation for uniformly distributed angles ($45^{\rm o}$). $68\%$, $90\%$ and $95\%$ contours combining bootstrap and PA errors are overlaid in orange shades. The pink dashed curve shows the $95\%$ contour obtained re-sampling PA errors only.  {\it Right panel}:  Dependence of $\langle \theta_{\rm low} \rangle$ versus $\langle \theta_{\rm high} \rangle$ on the varying mass threshold, as indicated by circles of varying orange shades from light ($10^{9.5} \,M_{\odot}$) to dark ($10^{11.2} \,M_{\odot}$). Results are for the SAMI sample with 3 $\sigma$ GAMA filaments.}
\label{fig:angle2}
\end{figure*}

Fig~\ref{fig:angle2}, left panel displays as a black circle the average angle in the low-mass subsample $\langle \theta_{\rm low} \rangle$ versus the average angle in the high-mass subsample $\langle \theta_{\rm high} \rangle$ calculated using Method 1 for redshift distortion correction (black circle) and using $M_{\rm thresh}=10.8$ which maximises signal-to-noise while still maintaining both the low and the high mass signals clearly visible, i.e. above than the marginalized $1-\sigma$ thresholds in each direction. Red-orange contours are computed combining bootstrap to PA errors, assuming gaussian errors on position angles as previously described. Darker to lighter shades indicate the regions in which $50 \%$, $68 \%$  and $90 \%$  of such signals lie. The uncertainties on the signal are completely dominated by the low statistics, with only a small contribution from position angle errors (see the pink dashed contour, which shows the $95\%$ contour obtained re-sampling PA errors only for the full sample).
Vertical and horizontal black dashed lines show the expectations in each sample for uniformly distributed angles ($45^{\rm o}$). 
 Blue shaded areas and dashed contours show the distribution of the spurious noise obtained using the first method described above. While substantial, the noise show no bias and is centred on $\langle \theta_{\rm low} \rangle=\langle \theta_{\rm high} \rangle=45^{\rm o}$.  
 
As suggested by  Fig.~\ref{fig:align1}, the average angle  $\langle \theta_{\rm low} \rangle=43.6\pm0.7^{\rm o}$  in the low-mass sub-sample extracted from the full sample is significantly lower than the uniform expectation, implying a degree of alignment of stellar spins with  their nearby filament. Conversely, the average angle  $\langle \theta_{\rm high} \rangle=46.5\pm1.5^{\rm o}$  in the high-mass sample is significantly higher, revealing a tendency to display a spin orthogonal to nearby filaments in this sample. It is remarkable that, despite the low number of galaxies in our sample (1418) and the expected faintness of such signals at $z<0.1$, both the low-mass and the high-mass signals are recovered above the $95\%$ confidence interval, with a typical probability that such a pair of signals in the correct quadrant be spurious  $<1\%$. Note that this probability even decreases to $<0.03\%$ when increasing the mass threshold above $10^{11}\,M_{\odot}$, although it is at the expense of the low-mass alignment signal. Indeed, further increasing the mass threshold allows to recover the pair of signals with a level of confidence $>\rm 2 \sigma$ but this is mostly a confirmation of the high-mass orthogonality signal as the low-mass signal is faint in this case (see Appendix.~\ref{section:refine} for a similar effect with the $2\sigma$ filaments). This provides strong motivation to work towards larger statistical samples of IFS galaxies such as the one to be provided by the upcoming Hector survey \citep{BlandHawthorn11,Bryant16}. 
 
We also varied the mass threshold used to segment the SAMI sample into a low-mass and a high-mass sub-samples. In Fig~\ref{fig:angle2}, right panel, circles in shades of light yellow to dark orange indicate the average point obtained for a variety of mass thresholds. One can see that increasing the mass threshold from $10^{9.5}\,M_{\odot}$ to $10^{11.2}\,M_{\odot}$ progressively decreases the average spin-filament angle in the lower mass sample while that of the higher mass sample increases. This is consistent with the fact that progressively adding more and more massive galaxies to the lower mass sample decreases the average tendency of galaxies in this sub-sample to align their kinematic spin axis with cosmic filaments. On the other hand, limiting the higher mass sample to more and more massive galaxies highlights a stronger orthogonal orientation of the spin with respect to filaments. The two signals are jointly detected above $10^{10.4} \, M_{\odot}$. This confirms the existence of a transition stellar mass in the  $10^{10.25}-10^{10.75} \, M_{\odot}$ ranges as suggested in \cite{Codis15} and \cite {Codis18} using Horizon-AGN, although with a differently calibrated filament reconstruction. 

\begin{figure}
\center \includegraphics[width=0.95\columnwidth]{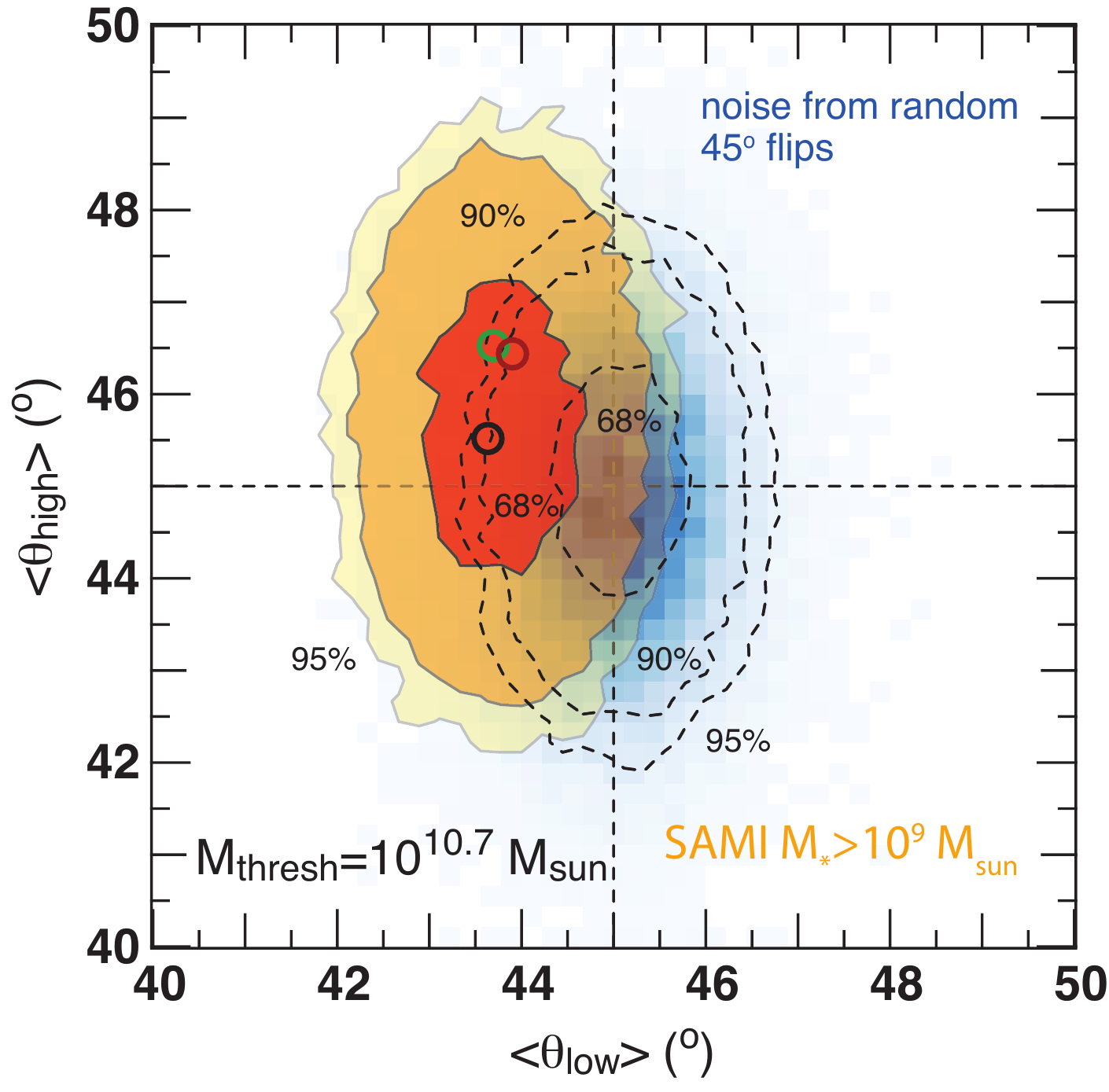}
  \caption{Average spin-filament angle for galaxies with $M_{*}< 10^{10.7}\, M_{\odot}$ (``low") versus average angle for galaxies with $M_{*}> 10^{10.7}\, M_{\odot}$ ``high") using three redshift distortion correction methods (black, red and green circles). Shades of blue and dashed contours indicate the distribution of values for the expected level of noise, obtained from random $45^{\rm o}$ flips of galaxy kinematic axis. Straight black dashed lines show the expectation for uniformly distributed angles ($45^{\rm o}$). Solid black contours show the uncertainties combining bootstrap and gaussian errors on position angles.}
\label{fig:angle3}
\end{figure}

Similarly, to assess the importance of the various methods used and parameters chosen, Fig~\ref{fig:angle3} displays the average angle in the low-mass sub-sample $\langle \theta_{\rm low} \rangle$ and the average angle in the high-mass subsample $\langle \theta_{\rm high} \rangle$ and for the three distinct methods used to correct for redshift space distortions (respectively red, green and black circles for methods 0, 1 and 2).  We also use a slightly lower mass threshold $M_{\rm thresh}=10^{10.7}\, M_{\odot}$ to illustrate the typical evolution of signals with the mass threshold. Here we use the second method to estimate the level of noise (random flips by $45^{\rm o}$). This makes no significant difference on the contours. As in the previous plots, blue shaded area and dashed contours show the distribution obtained for the noise, this time flipping the kinematic axis by 
$45^{\rm o}$ or $-45^{\rm o}$. Orange contours show the error on the signal combining bootstrap and position angle errors. 

Fig~\ref{fig:angle3} shows that the signal persists when the mass threshold is decreased, but as expected with a small decrease of the high-mass orthogonal signal. The choice of redshift distortion corrections is limited to a random walk along the y-axis in the upper half of the $1-\sigma$ region which does not impact the significance of the detection. Expectedly only the high-mass signal is noticeably impacted as most galaxies in massive groups that are subject to the strongest redshift distortion effects are found in this mass range.

Overall, in all cases presented here, our analysis finds a clear signature of mass dependent spin flips for galaxies in the SAMI sample. In the following section, we analyze in details the impact of the mass threshold used to define these samples.

\subsection{Effect of mass threshold}
\label{section:mass}

\begin{figure}
\center \includegraphics[width=0.95\columnwidth]{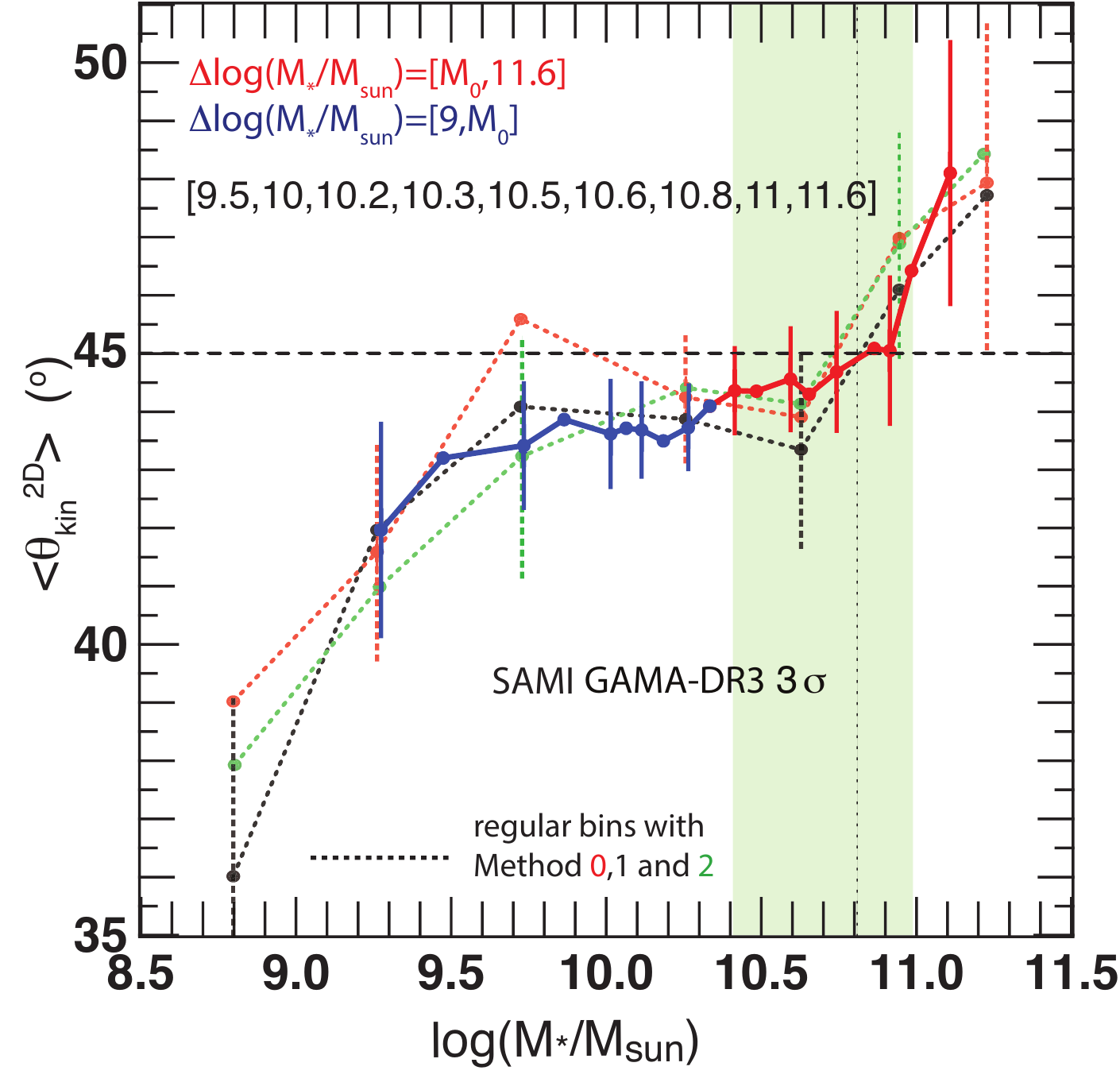}
  \caption{Evolution of $\langle \theta_{\rm kin}^{\rm 2D} \rangle$ (in each bin) as a function of the median mass in the bin $\delta M_{*}$ in SAMI using standard (dotted lines) and irregular shifting (solid line) bins. In the latter case, low-mass bins are plotted in blue while high-mass bins appear in red.$\langle \theta_{\rm kin}^{\rm 2D} \rangle$ increases steadily with median stellar mass. The horizontal dashed line shows the expectation for uniformly distributed angles. The green shaded area shows the range in which the transition mass can be bracketed.}
\label{fig:mass-th-sami}
\end{figure}

Let us now estimate the dependence of $\langle \theta_{\rm kin}^{\rm 2D} \rangle$ on stellar mass. Fig.~\ref{fig:mass-th-sami} plots $\langle \theta_{\rm kin}^{\rm 2D} \rangle$ as a function of median stellar mass in each bin $\delta M_{*}$ or the full SAMI sample. The black dashed line shows the expectation for random (uniformly distributed) orientations of spins. Dotted red, black and green curves display the evolution using standard consecutive, non-overlapping independent mass bins, using redshift distortion correction method 0, 1 and 2 respectively. One can see that $\langle \theta_{\rm kin}^{\rm 2D} \rangle$ displays a progressive increase with stellar mass, from values well below the expectation from randomly distributed angles ($45^{\rm o}$) at low mass to values above it at large masses. Note that in this particular case, we added an extreme low-mass bin containing the 52 galaxies with $10^{8.5}\,M_{\odot}<M_{*}<10^{9}\,M_{\odot}$ that pass all other selection cuts. It confirms that the trend extends to this mass range, although we exclude it in the rest of the study to avoid contamination of our results by low-reliability PAs.

 These results highlights a progressive transition from an aligned to an orthogonal orientation of the spin axis with respect to the nearby cosmic filament as stellar mass increases.
Nonetheless, a significant level of noise is visible on the signal, especially in the case where we simply do not correct for redshift distortions (Method 0, in dotted red). It is expected as in this case, many galaxies in massive groups are wrongly assigned to a spurious filament arising from finger-of-god effects, especially in the outskirts of such groups. Unsurprisingly, the smoothest signal is obtained with Method 2 (green), which simply excludes such galaxies hence puts a strong focus on filament galaxies. Method 1 (black) shows however very consistent results with the advantage of extending results to all group galaxies with only limited misidentifications.

To overcome low number statistics and smooth out local environmental/redshift distortion effects, we stick to Method 1 but define irregular, overlapping mass bins, making sure that the median stellar mass (in the bin) smoothly and steadily increases across consecutive bins. To do so we apply a binning procedure similar to what was done in Fig.~\ref{fig:mdfil}. We define:
\begin{itemize}
\item ``Blue" bins: each bin contains all galaxies with a stellar mass between $10^{9.5}\,M_{\odot}$ and $10^{M_{\rm i}}\,M_{\odot}$ with $M_{\rm i}$ taking values 10, 10.2, 10.3, 10.5, 10.6, 10.8, 11 and 11.6.
\item ``Red" bins: each bin contains all galaxies with a stellar mass between $10^{M_{\rm i}}$ and $10^{11.6}\,M_{\odot}$ with $M_{\rm i}$ taking values 9.5, 10, 10.2, 10.3, 10.5, 10.6, 10.8 and 11.
\end{itemize}

One can see that $\langle \theta_{\rm kin}^{\rm 2D} \rangle$ now increases steadily, monotonously and smoothly with stellar mass at masses below $10^{10.4}\,M_{\odot}$ to values above it at masses above $10^{10.9}\,M_{\odot}$. Similar results are obtained with Method 0 and 2. This procedure effectively smoothes out the curve (integration-like procedure), therefore allowing us to identify more precisely the transition mass as the robust crossing of the $\langle \theta_{\rm kin}^{\rm 2D} \rangle=45^{o}$ line, and the range around it where $\langle \theta_{\rm kin}^{\rm 2D} \rangle$ is compatible with $45^{o}$ within $1\sigma$ error bars.

 The transition mass characterizing the disappearance of the alignment signal and the onset of the orthogonal trend can therefore be confidently bracketed between $10^{10.4}\,M_{\odot}$ and $10^{10.9}\,M_{\odot}$. This trend is consistent with previous results: low-mass galaxies tend to display $\langle \theta_{\rm kin}^{\rm 2D} \rangle<45^{\rm o}$ , indicating a degree of spin alignment to nearby filament while high mass galaxies samples have $\langle \theta_{\rm kin}^{\rm 2D} \rangle >45^{\rm o}$, indicating a tendency of their spin to lie orthogonally to the nearby cosmic filament. 
\section{Comparison to simulations}
\label{section:sims}

Using the SAMI galaxy sample, we found a clear signature for galaxies to transition progressively from having a spin aligned with their nearby cosmic filament at low stellar mass to a spin orthogonal to it at high stellar mass. In the following section, we analyse the corresponding signals in two different types of simulations: the Horizon-AGN, a state-of-the-art cosmological hydrodynamic simulation described in Section.~\ref{section:obs} and a GAMA mock lightcone produced with the semi-analytical model Shark \citep{Lagos18} then passed through the lightcone software Stingray (Obreschkow et al., in prep).

\subsection{The hydrodynamic simulation Horizon-AGN.}
\label{section:hagn}

Let us focus on Horizon-AGN, as described in Section~\ref{section:obs}. Recall that we extracted two types of filamentary network in Horizon-AGN: the {\it galaxy} filaments, obtained directly from the distribution of galaxies and the {\it gas} filaments, obtained from the gas density on cosmic scales in the simulation. In the following analysis, we present results for three different samples:
\begin{itemize}
\item The full Horizon-AGN galaxy population with $M_{*}>10^{9.5}\,M_{\odot}$, in combination with the gas filaments. The main advantages of this selection are the more realistic stellar mass function and a cosmic extraction that directly traces the density filaments, i.e. the underlying environment that gave rise to the population of galaxies. This cosmic web extraction is also independent of the galaxy population itself.
\item A SAMI mock population resampling the Horizon-AGN population to match the SAMI stellar mass function (in mass bins of 0.25 dex), in combination with either gas or galaxy filaments, extracted directly from the distribution of galaxies. This measure is conceptually more comparable to SAMI, especially when using galaxy filaments. An important caveat is that galaxies with $10^{9}\,M_{\odot}<M_{*}<10^{9.75}\,M_{\odot}$ were replaced by galaxies with $10^{9.5}\,M_{\odot}<M_{*}<10^{9.75}\,M_{\odot}$ (randomly drawn) in Horizon-AGN as lower masses galaxies lack resolution to allow a reliable computation of their angular momentum.
\end{itemize}
 We also checked that excluding galaxies within one virial radius of clusters with halo masses $10^{14}\,M_{\odot}$ to account for the fact that the GAMA region across which SAMI galaxies are found contains no such clusters made no significant difference.

Following Fig.~\ref{fig:angle2}, Fig.~\ref{fig:thetahagn} displays the average angle in the low-mass subsample $\langle \theta_{\rm low} \rangle$ versus the average angle in the high-mass subsample $\langle \theta_{\rm high} \rangle$ calculated for the full Horizon-AGN population and gas filaments on the left panel, for the SAMI mock population and gas filaments on the middle panel and for the SAMI mock population and galaxy filaments on the right panel. Straight black dashed lines show the expectations in each sample for uniformly distributed angles ($45^{\rm o}$). The black open circle and orange contours show the results for the Horizon-AGN sample. We varied the mass threshold as we did for SAMI and present results obtained for the mass threshold that maximises signal-to-noise while maintaining each individual signal above the marginalized $1-\sigma$ threshold: $M_{\rm thresh}=10^{10.7}\, M_{\odot}$. Red-orange contours are computed using bootstrap. Darker to lighter shades indicate the regions in which $50 \%$, $68 \%$  and $90 \%$  of such signals lie.  In addition, various SAMI detections and corresponding 1-$\sigma$ contours are overlaid in pink, green and purple for comparison.

\begin{figure*}
\center \includegraphics[width=2\columnwidth]{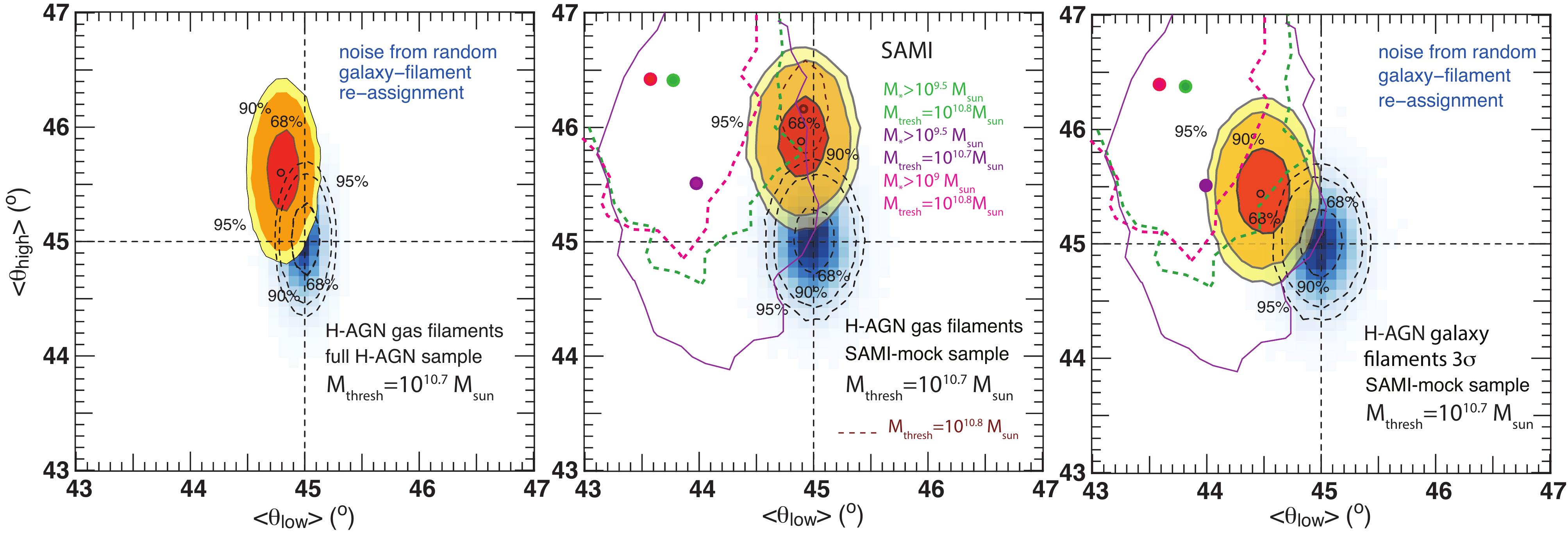}
  \caption{Average spin-filament angle for galaxies with $M_{*}< M_{\rm thresh}$ ("low") versus average angle for galaxies with $M_{*}> M_{\rm thresh}$ ("high") in the Horizon-AGN simulation. Shades of blue and dashed contours indicate the distribution of values for the expected level of noise, obtained from random re-pairing of galaxies and filaments. Straight black dashed lines show the expectation for uniformly distributed angles ($45^{\rm o}$). Solid black contours show the  bootstrap errors. Orange contours show the signal obtained for:{\it Left panel:}  the full Horizon-AGN galaxy population with $M_{*}< 10^{9.5}\, M_{\odot}$ with gas filaments. {\it Intermediate panel:} the SAMI mock population with gas filaments. {\it Right panel:} the SAMI mock population with $3\sigma$ galaxy filaments. SAMI signals with specific sub-samples and mass thresholds are overlaid with 1-$\sigma$ contours in pink, green and purple.}
\label{fig:thetahagn}
\end{figure*}

In all cases, a signal qualitatively similar to that found is SAMI is recovered, with more massive galaxies showing a tendency to orientate their spin orthogonal to their nearest cosmic filament (average angle   $\langle \theta_{\rm high} \rangle>45^{\rm o}$) while their less massive counterparts are more likely to display a parallel orientation ($\langle \theta_{\rm low} \rangle<45^{\rm o}$). The robustness of the detection, above the $2\sigma$ level and close to the $3\sigma$ level when using the galaxy filaments, is also slightly stronger than that in SAMI. The optimal transition masses (selected similarly to what was done in SAMI) are however similar in all cases. It is nonetheless important to notice that both the signal amplitude and the level of noise in Horizon-AGN at $z=0$ are smaller than those presented in SAMI. The average angle in the low mass sample are respectively $\langle \theta_{\rm low} \rangle=44.76^{\rm o}$, $\langle \theta_{\rm low} \rangle=44.85^{\rm o}$ and $\langle \theta_{\rm low} \rangle=44.45^{\rm o}$ from left to right, while the average in the high mass sample are respectively $\langle \theta_{\rm high} \rangle=45.6^{\rm o}$, $\langle \theta_{\rm high} \rangle=45.85^{\rm o}$ and $\langle \theta_{\rm high} \rangle=45.4^{\rm o}$. Several considerations can explain these variations:
\begin{itemize}
\item The uncertainties are much higher in the SAMI sample compared to simulated sample due mostly to lower numbers (and with a contribution from errors on redshifts and position angles). Indeed, the typical error $<0.3^{\rm o}$ in the case of simulated samples typically increases to $\approx 1.5^{\rm o}$ in the case of the SAMI sample. Hence while the amplitude of the signal seems stronger in SAMI, it is in fact not more significant than in Horizon-AGN. Within error bars, the high mass signals are actually compatible with one another but a tension of $0.3^{\rm o}$ to $0.5^{\rm o}$ still exists for the low mass samples.
\item An important contribution of the low-mass SAMI signal comes from galaxies with $M_{*}<10^{9.5}\,M_{\odot}$ which are either not resolved or under-resolved in Horizon-AGN hence could not be used for this analysis. Therefore we expect a strong decrease of the alignment signal for low-mass sample in the simulation as the important contribution for lowest mass galaxies is absent. Removing galaxies with $M_{*}<10^{9.5}\,M_{\odot}$ from the SAMI sample (green and purple dots and contours) decreases the discrepancy in the low-mass signal, especially on the right panel i.e when using both the SAMI mock sample and the galaxy filaments for Horizon-AGN, which is the most comparable to the prodecure used in SAMI.
\item Moreover, even above this stellar mass, \cite{VdS18} confirmed quantitatively that the formation of thin discs with high $v/\sigma$ is hindered in Horizon-AGN, which previous works \citep{Dubois14, Welker17} attribute in part to spatial resolution (which also limits the resolution of numerous physical processes from star formation to AGN feedback). Discs therefore appear very puffed-up with minimal minor-to-major axis ratios of order 0.2. This persists across a wide range of stellar masses and increases stochasticity on the orientation of their angular momentum, especially when computed in one half-luminosity radius compared to galaxies observed with SAMI. Interestingly, focusing on the 1278 best resolved SAMI galaxies, which effectively trims the sample of its smallest galaxies, mostly with $M_{*}<10^{10.3}\,M_{\odot}$, removes the discrepancy of  low-mass signals between SAMI and Horizon-AGN (see Appendix.~\ref{section:spatial}).
\item The galaxy mass function in Horizon-AGN is a reasonable fit to that derived from observations but still overshoots it by a factor $\approx 3$ at the low ($M_{*}<10^{9.5}\,M_{\odot}$) and high mass ends ($M_{*}>10^{11}\,M_{\odot}$) \citep{Kaviraj17,Canas18}, implying that the average stellar-to-halo mass ratio is not as well matched to real galaxies in these mass ranges, which might artificially reduce the alignment effect inherited from the halo. Reducing the SAMI optimal mass threshold ($10^{10.8}\, M_{\odot}$) to the one found in Horizon-AGN ($10^{10.7}\, M_{\odot}$) removes the tension between the true and simulated high-mass signals (purple dot and contours, see also Appendix.~\ref{section:spatial}).
\end{itemize}

Despite minor differences, the simulated and observed signals are fully consistent with one another. As could be expected, the low-mass signal obtained with the galaxy filaments is increased (by a factor 2) compared to the one obtained with the gas filaments. This is consistent with the idea that reconstructing the cosmic web from the galaxy distribution allows to recover a number of galaxy tendrils around the biggest filament, even with a relatively high signal-to-noise cut since such structures are topologically robust in the galaxy distribution although much fainter at $z=0$ in the gas density field. Mostly low-mass galaxies are found around these smaller scales, closer and more laminar, filaments and the alignment signal around them is therefore strengthened.

Our results are also broadly consistent with the recent study in Horizon-AGN by \cite{Codis18}, with the exception of a few expected caveats detailed hereafter. It should be noted that their study is carried out in full 3D and not in projection, and on the total stellar angular momentum of galaxies rather than the angular momentum in one effective radius. Other differences include the exact persistence thresholds used to recover the cosmic web (hence the scales recovered) and the way the alignment is estimated. Their results are comparable to ours in terms of trend with stellar mass: the spin is orthogonal at high stellar mass but this trend fades away at lower mass, with a transition mass consistent with ours. However they indicate that the low-mass alignment trend is hardly visible on the $\xi$ distribution at $z=0$ (while visible at higher redshifts). Consistently, we do find that the amplitude of the low-mass signal is lower and less significant than that of the high mass signal when using gas filaments, but we do not find that it is compatible with a non-detection at $z=0$. This actually comes from the fact that they do not quantify the alignments trends in terms of average spin-filament angles. Doing so with the 3D Horizon-AGN angular momentum catalogs they use actually allows to recover the spin alignments at low-mass at $z=0$, although it is faint. Moreover, low-mass galaxy spins are expected to align not only with their nearest filament but also within their nearby wall. As a consequence, observing the signal in projection can enhance the signal as it suppresses the contribution to the misalignment along the line of sight and all the more that in this case, it can result of a mix of the two alignment signals. It is also important to stress that \cite{Codis18} chose to include all galaxies with $M_{*}>10^{8.5}\,M_{\odot}$ in the comparison, while galaxies with $M_{*}<10^{9.5}\,M_{\odot}$ do not have well-resolved kinematics comparable to those found in observations and add to the noise.

\begin{figure*}
\center \includegraphics[width=1.99\columnwidth]{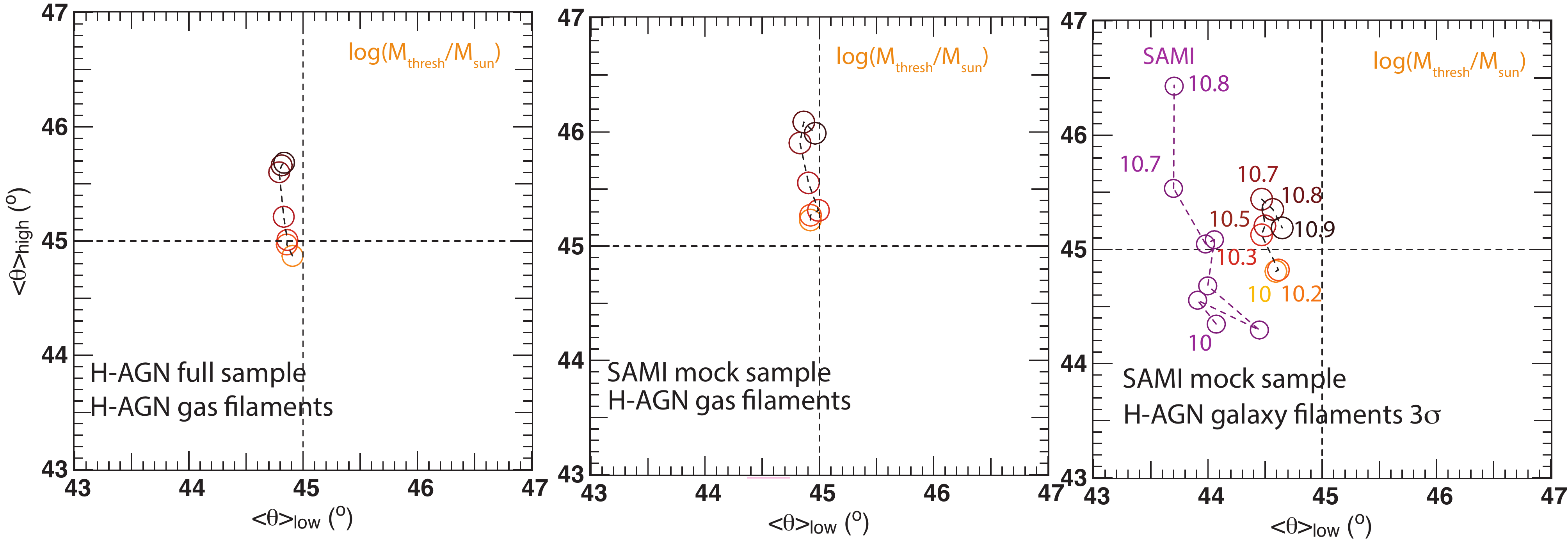}
  \caption{Dependence of $\langle \theta_{\rm low} \rangle$ versus $\langle \theta_{\rm high} \rangle$ on varying mass thresholds, as indicated by circles of varying orange shades from light ($10^{10}\,M_{\odot}$) to dark ($10^{10.9}\,M_{\odot}$). {\it Left panel}: full Horizon-AGN sample with gas filaments. {\it Intermediate panel}: SAMI-mock galaxy sample with gas filaments. {\it Right panel}: SAMI-mock galaxy sample with $3\sigma$ galaxy filaments.}
\label{fig:mass-th}
\end{figure*}

They also account for all star particles in their computation of spins and therefore include an important component from the outskirts which could have been formed only recently in a cosmic environment where gas is scarce and vorticity quadrants highly turbulent. Indeed, most of the filament alignment is expected to be acquired at high redshift when diffuse accretion dominates the mass intake \citep{Dubois14,Laigle15,Codis18}. In addition, properly segmenting galaxy outskirts in simulations is a difficult task that most galaxy finders only approximate \citep{Canas18}. Being at large radii, corresponding stellar particles can nonetheless skew the angular momentum measurement.

\subsection{Effect of mass threshold in Horizon-AGN.}

Fig.~\ref{fig:mass-th} displays the evolution of the relation between $\langle \theta_{\rm low} \rangle$ and $\langle \theta_{\rm high} \rangle$ when increasing the mass threshold used to segment SAMI galaxies into a lower and a higher mass sub-samples for the three different Horizon-AGN samples. Both the lower and higher mass sample signals are recovered for a wide range of stellar mass thresholds. Focusing on the rightmost panel (SAMI mock sample with the galaxy filaments) which is the most comparable to SAMI, both signals are recovered for threshold masses  $>10^{10.3}\,M_{\odot}$, similarly to what is seen in SAMI. The trend for more massive galaxies to flip their spin orthogonal to their nearest filament at higher mass threshold is expected but the limited variation seen at low-mass is more surprising. It is actually an effect of the mass function, biased towards lower masses.  As a consequence, increasing the mass threshold does not impact much the statistics in the low mass sample.

The higher mass signal increases with mass threshold up to $10^{10.7}\,M_{\odot}$ consistently with previous results but it shows a decrease past this threshold mass. The significance of this decrement is limited due to the large increase in typical errors for the higher mass sample at high mass thresholds. It may also be related to the fact that most massive galaxies near the centre of groups are much more likely to be connected to multiple filaments when finer, more numerous galaxy filaments are used. This renders the assignment to a single one (here, the closest one but not always the most contrasted one) less reliable.
This is especially true in Horizon-AGN that host too many galaxies with $M_{*}<10^{9.5}\,M_{\odot}$ compared to GAMA (a combination of the mass function bias in the simulation and the detection bias in the survey), susceptible to contribute importantly to such filaments. Note also that, in GAMA, our redshift distortion correction further limits the possibility for galaxies in a massive group to form local, lower contrast filaments, as those are often stretched along the line of sight and therefore trimmed from the skeleton. This suppression of the orthogonal trend for most massive galaxies in the simulation is the cause of the tension between the simulated and observed high mass signals. Horizon-AGN also hosts too many high mass galaxies which might also contribute to the discrepancy. 

Nonetheless in this case both signals are detected simultaneously across a large range of threshold masses considered, compatible with observed trends in SAMI. To better constrain the transition mass, Appendix.~\ref{section:evolm} reproduces the evolution of the average spin-filament angle with stellar mass in Horizon-AGN. From this comparison, it becomes even clearer that the tensions between Horizon-AGN and SAMI arise from the extreme parts of the mass range considered in this study: $ M_{*}<10^{9.5}\,M_{\odot}$ and $ M_{*}>10^{11}\,M_{\odot}$, i.e. from the less reliable populations in the simulation. 

This consistency between SAMI galaxies and simulated galaxies otherwise strongly supports the  scenario where low-mass galaxies build up their angular momentum parallel to their nearby filament which have aligned vorticity, while that of more massive galaxies is dominated by mergers along filaments. However, while this scenario was first robustly established for dark matter haloes \citep{Codis12} and later extended to galaxies \citep{Dubois14,Codis18}, how it is transferred to galaxies is still debated. How much is purely inherited from the host halo and how much can be attributed to the specificity of galaxy mergers (and in particular the anisotropic distribution of their orbits with respect to the cosmic web) or to collimated gas streams? More generally, do anisotropic hydrodynamic processes play a major role in amplifying this connection or is it fully accounted for considering the tidal influence of the host halo on its galaxy and smooth accretion onto it? To get a clearer idea of the mechanisms at stake, in the next section we reproduce our analysis using a mock GAMA lightcone computed from the semi-analytical model Shark \cite{Lagos18}, applied to pure N-body simulations.

\subsection{Semi-analytical model SHARK.}
\label{section:shark}

The lightcone presented in this section is obtained after applying the Shark semi-analytical model developed by \cite{Lagos18} on the largest dark matter simulation of the SURFS suite \citep{Elahi18} and using it as an input for the mock lightcone generator Stingray (Obreschkow et al, in prep). The cubic, 210 Mpc $\rm h^{-1}$ on a side SURFS simulation contains $1504^{3}$ dark matter particles, hence resolves dark matter haloes down to $\approx 5\times10^{9}\,M_{\odot}$. Haloes and sub-haloes are identified with the phase-space structure finder VelociRaptor \citep{Elahi18} for each snapshot and merger trees are produced with the TreeFrog software \citep{Poulton18}. The Shark model is then applied and galaxies are formed and evolved through cosmic time using as backbone this population of haloes. Shark is a modular software allowing to combine a wealth of physical models for a variety of processes into a self-consistent semi-analytical model. In this study, we focus on the flagship Shark configuration, which includes elaborate physics recipes for cooling, photo-ionization, star formation, stellar and AGN feedback, chemical evolution, bulge and disc formation, galaxy mergers and disc instabilities. Details can be found in  \cite{Lagos18}. A key component of this model is that, similarly to all existing semi-analytical models, the behaviour of the gas (and therefore of the stars) is derived directly from that of the halo through effective analytical prescriptions. In particular, the angular momentum of gas and stars is directly derived from that of the halo through radially integrated (and therefore isotropic) recipes, therefore not capturing most of the specific effects of the hydrodynamics. Note that in our model, only the orientation of the gas spin is derived from that of the host halo, the spin parameters are assigned randomly. In the case where satellites cannot be assigned to a sub-halo anymore (after fading away through dynamical friction for instance), the orientation of the angular momentum is also drawn randomly. However, this latter case is unfrequent in our region of interest in the GAMA mock lightcone which contains no cluster, and similar satellites show little to no significant spin alignment in hydrodynamical simulations either \citep{Codis18}.

\begin{figure*}
\center \includegraphics[width=2\columnwidth]{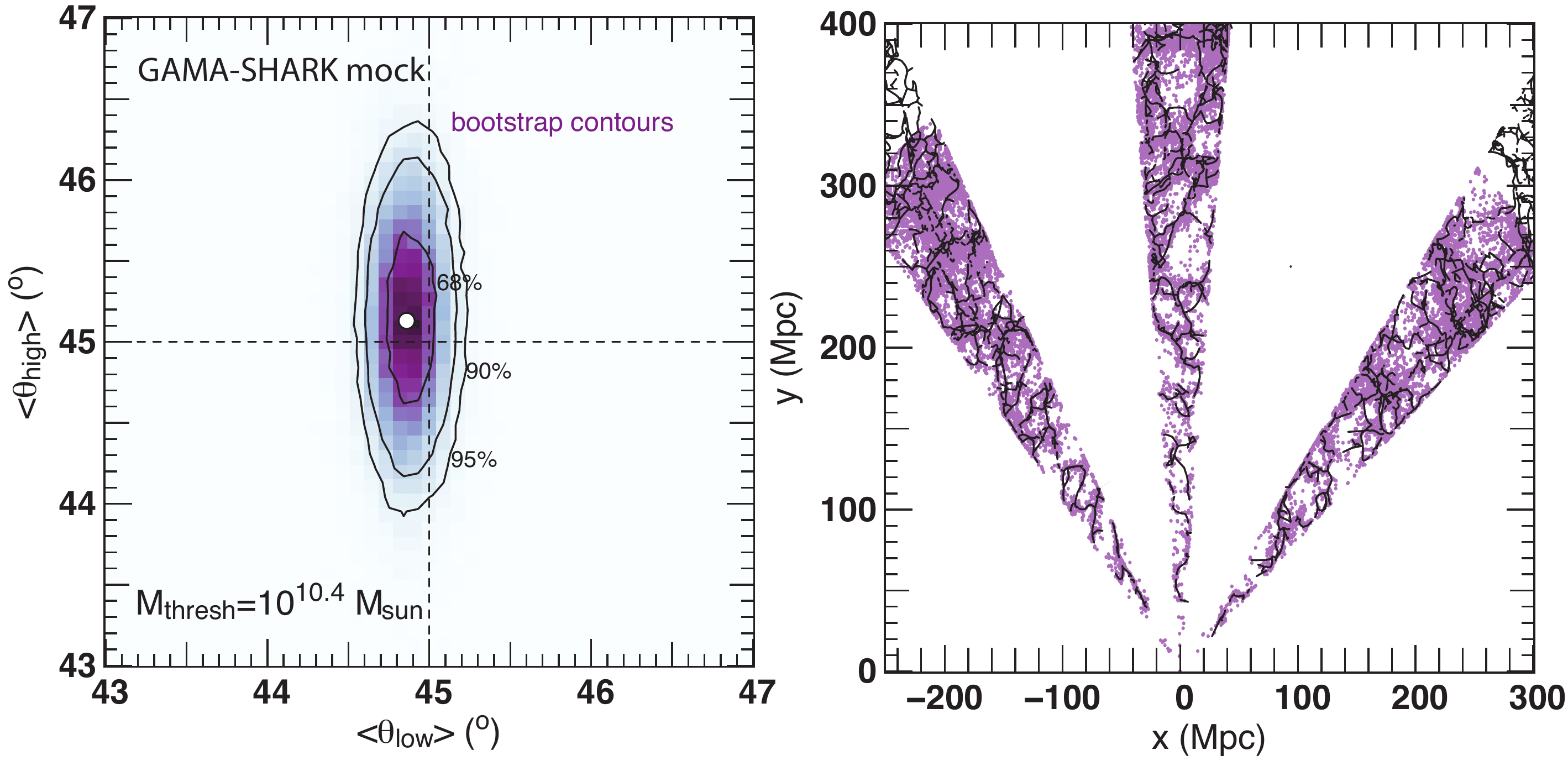}
  \caption{{\it Left panel:} Average spin-filament angle for galaxies with $M_{*}< M_{\rm thresh}$ ("low") versus average angle for galaxies with $M_{*}> M_{\rm thresh}$ ("high") in the Shark GAMA mock lightcone.Contours show the bootstrap errors {\it Right panel}: Projected map of the cosmic filaments (black solid lines) in the mock GAMA lightcone, on top of the mock galaxy distribution itself (purple dots).}
\label{fig:shark}
\end{figure*}

 Fig.~\ref{fig:shark}, right panel, shows the reconstruction of the $3\sigma$ cosmic web (solid black lines) over the mock GAMA galaxy distribution (purple dots) for mock fields G09, G12 and G15 for $z<0.1$ using all galaxies with $M_{*}>10^{8.5}\,M_{\odot}$. The procedure used in SAMI is repeated here: galaxies with $M_{\rm *}>10^{9} \,M_{\odot}$ are assigned to their nearest filament in 3D (real space). Then the difference between mock spin position angles (2D projected spin axis) and the position angle of filaments is computed. Fig.~\ref{fig:shark}, left panel shows the average angle in the low-mass subsample $\langle \theta_{\rm low} \rangle$ versus the average angle in the high-mass subsample $\langle \theta_{\rm high} \rangle$ for the mass threshold that maximises signal-to-noise : $M_{\rm thresh}=10^{10.4}\, M_{\odot}$ with $68\%$, $90\%$ and $95\%$ contours obtained from bootstrap.
 
 While the signal is qualitatively consistent with what is found in SAMI and Horizon-AGN, the significance of it is much smaller. Despite the high statistics in the sample, the uncertainties on $\langle \theta_{\rm 2D}^{\rm kin} \rangle$ are higher than in Horizon-AGN while the high mass signal is weaker (average angle of $45.15^{\rm o}$), even marginally compatible with a non-detection (the $1\sigma$ contour crosses the $45^{\rm o}$ lines). The low-mass alignment signal is more significant and comparable to the one found in Horizon-AGN with gas filaments (average angle of $44.8^{\rm o}$) but much smaller than the signal found for SAMI galaxies and the galaxy filaments in Horizon-AGN likewise while this sample does not suffer from the low-mass limitations of hydrodynamic cosmological simulations. 
 
 Three important conclusions can be drawn from this result:
 \begin{itemize}
 \item A faint signature of spin flips is recovered, emphasizing the transfer of such trends from the large tides traced by host dark matter haloes.
 \item The signal is very faint, suggesting that a proper treatment of hydrodynamic processes is necessary to properly reflect the transfer of the spin transition from haloes to galaxies. This includes taking into account the fact that gas shocking and cooling in the cosmic web leads to more collimated filaments than those derived from dark matter only \citep{Pichon11}. In particular, since galaxies are forming in these flows, the distribution of satellites and pre-merger orbits is more collimated for galaxies than for dark matter sub-haloes. This should be taken into account when populating haloes in semi-analytical models. More generally, accounting for the increased collimation of gas flows would improve such models to better constrain accretion onto galaxies.
 \item Let's emphasize that in pure DM simulations, the halo spin transition signal with halo mass is strong \citep{Porciani02,Aubert04,Aragon07,Hahn07,Paz08,Bett12,Codis12}. Therefore our results highlights the fact that, although arising from similar processes (vorticity, mergers), galaxy spin alignment transition with stellar mass is not merely a direct inheritance of the haloes' spin behaviour (connected through the halo-to-stellar mass correlation) and certainly not synchronized to it. In our model, absence of consideration for the degree of collimation the gas adds to the galaxy distribution, improper treatment of galaxy spins in galaxy mergers (the remnant still merely follows the spin orientation of the DM halo) and of satellite galaxy spins in low-resolution (still align to the subhalo's spin) or unresolved subhaloes (random spin orientation) is sufficient for the transition with stellar mass to fade away.
 \end{itemize}

\subsection{Comparison with past observations.}
\label{section:discuss}

Our results are consistent with observations in the SDSS using shape as a proxy for spin \citep{Tempel13a,Tempel13b,Pahwa16} and with \cite{Chen19} which focuses on the galactic plane orientation of massive galaxies. However, they differ from the more directly comparable study performed on 2736 galaxies in the IFS survey MaNGA \citep{Krowleski19}. In this study, the authors do not find any significant alignment for their full sample, which is expected given their mass function which samples stellar masses above and below the expected transition mass. However, their analysis of the mass dependence of the alignment is partially inconsistent with our results and with the simulations they compare to. While they may find a small tendency to spin alignment for galaxies $M_{\rm *}<10^{10} \,M_{\odot}$, they find a similar amplitude alignment signal for $M_{\rm *}>10^{11} \,M_{\odot}$, at odds with the expected orthogonal orientation which is also expected to be the most robust of the two trends \citep{Pahwa16,Chen19}.

While this suggests that larger samples will be necessary to reach a definite conclusion, a number of issues might also impact their measurements. In particular, in \cite{Krowleski19}, the method to extract the orientation of cosmic filaments might be at stake. It is indeed purely two-dimensional (resulting in a complete mitigation of walls and filaments, with a sample likely dominated by walls), relies heavily on smoothing and limits the orientation estimation to the ten nearest galaxies. It is in particular worth noticing that the same method fails to detect any significant signal in the Illustris simulation where the 3D signal is however established by the same study \citep{Krowleski19}. This highlights the importance of choosing a filament extraction algorithm with a strong focus on recovering the spine of filaments and walls with their precise orientation (i.e. the local structure of the cosmic web), albeit at the expense of its abiltity to define``thickness" or ``extent" parameters for such structures, or to assign galaxies to one type of structure.
Other differences that might impact the results are the wider IFS field of view which gives weight to the outskirts of galaxies, more sensitive to their immediate environment,  and a very restrictive cut on PA errors which would lead to discard many well converged fits for slow rotators, therefore disproportionately biasing the high-mass sample towards its fastest rotators, likely to display the least orthogonal spins \citep{Codis18}. Finally the use of $\cos \theta_{\rm 2D}$ instead of $\theta_{\rm 2D}$ as the test variable complicates their analysis as its distribution (and corresponding poissonian errors on it) is strongly non-uniform for the null hypothesis (uniformly distributed 2D spins) hence makes the comparison subject to higher uncertainties for limited size statistical samples such as those used in IFS.

\section{Conclusion}
\label{section:conclusion}

We used the GAMA galaxy catalogs over the fields G09, G12 and G15 to reconstruct the density field on Mpc scales using a Delaunay tessellation. This density field allowed us to reconstruct the network of cosmic filaments around which SAMI galaxies were selected in the redshift range $0.01<z<0.1$. Using the kinematics of the 1418 SAMI galaxies found across those fields, we first verified that the  evolution of the average stellar mass with distance to filament is consistent with previous findings in simulations and observations: the average stellar mass increases closer to the spine of the filament, in agreement with the expected migration of galaxies towards these higher density regions as they grow in mass, and consistently with the higher accretion rate expected in such regions. We then explored the evolution of spin orientations in the frame of nearby cosmic filaments and compared our results with a hydrodynamic simulation and a semi-analytical lightcone. 

Our findings are:
\begin{itemize}
\item SAMI low-mass galaxies tend to display a spin aligned with the nearest cosmic filament while the spin orientation of their more massive counterparts is more likely to be orthogonal to their neighbouring cosmic filament. The stellar transition mass at $z<0.1$ is between $10^{10.4}\, M_{\odot}$ and  $10^{10.9}\, M_{\odot}$, consistently with predictions using hydrodynamical simulations from \cite{Dubois14} and \cite{Codis18}.
\item The same pair of signals is recovered in the full SAMI sample of GAMA galaxies as well as in the most resolved sub-sample with $M_{*}>10^{9.5} \, M_{\odot}$.
\item A lower amplitude signal of comparable significance is recovered in the Horizon-AGN simulation with well-matched transition masses for both the full galaxy population, using filaments extracted from the cosmic gas density field and for a SAMI mock with better matched stellar mass function, using filaments derived from the synthetic galaxy distribution.
 \item We find hints that the transition mass between the preferentially aligned and perpendicular orientations of galaxy spins varies with the mass scale used to define the filaments. Including more refined filaments seems to lead to lower transition masses.
 \item The similar analysis performed on a synthetic lightcone generated from a pure dark matter simulation with the Shark semi-analytical model reveals that the evolution of spin orientation in such models is qualitatively consistent with observations and hydrodynamical simulations but much reduced in amplitude.  While this supports some expected impact from dark matter tides, this also suggests that the enhanced anisotropy from baryon processes on large scales needs to be better accounted for in such models.
\end{itemize}

Although the number statistics in our sample is limited, these results provide strong motivation for the development of large scale IFS surveys sampling a wider variety of environments on a larger redshift range. In particular, recall that the present study was carried out in projection. Angles between filaments and galaxy spins are computed on the sphere from position angles. In deep but narrow surveys such as GAMA, this makes it hard to disentangle alignments with filaments and walls in close alignment. State-of-the art disc modelling combined with large statistics will be necessary to recover the 3D alignments of galactic spins with the zoology of cosmic structures, in particular filaments and walls. 

The low-mass signal is of particular interest since these alignments are attributed to the formation of quadrants of coherent vorticity (the curl of the velocity field) aligned with filaments in their vicinity, where young galaxies form, in hydrodynamics simulations and lagrangian theory likewise. This particular scenario predicts peaks of vorticity offset from filaments, dispatched on average in four quadrants of vorticity of alternate sign. If galactic spins indeed correlate spatially with these hydrodynamic structures, this suggests that galactic properties could be used to trace or even map out such cosmic flows providing that statistics is sufficient. Hints of such signatures in the SAMI sample will be investigated in an upcoming study (Welker et al., in prep).

\section*{acknowledgements}
{\sl 
The SAMI Galaxy Survey is based on observations made at the Anglo-Australian Telescope. The Sydney-AAO
Multi-object Integral-field spectrograph (SAMI) was developed jointly by the University of Sydney and the Australian Astronomical Observatory, and funded by ARC grants FF0776384 (Bland-Hawthorn) and LE130100198. JvdS is
funded under Bland-Hawthorn’s ARC Laureate Fellowship (FL140100278). 
JJB acknowledges support of an Australian Research Council Future Fellowship (FT180100231).
The SAMI input catalogue is based on data taken from the Sloan Digital Sky Survey, the GAMA Survey and the VST ATLAS Survey. The SAMI Galaxy Survey is supported by the Australian Research Council Centre of Excellence for All Sky Astrophysics in 3 Dimensions (ASTRO 3D), through project number CE170100013, the Australian Research Council Centre of Excellence for All-sky Astrophysics (CAASTRO), through project number CE110001020, and other participating institutions. The SAMI Galaxy Survey
website is http://sami-survey.org/ . 

This research was conducted via the Australian Research Council Centre of Excellence for All-sky Astrophysics in Three Dimensions (ASTRO 3D), through project number CE110001020.
This work was supported by the Flagship Allocation Scheme of the NCI National Facility at the ANU. This research was supported in part by the National Science Foundation under Grant No. NSF PHY-1748958.
Let us thank D.~Munro for freely distributing his {\sc \small  Yorick} programming language and opengl interface (available at \url{http://yorick.sourceforge.net/}). 
We warmly thank Sandrine Codis, Katarina Kraljic and Dan Taranu for valuable discussions during the preparation of this manuscript. We also thank S. Rouberol for running the Horizon cluster on which the simulation was post-processed
}
\vspace{-0.5cm}

\bibliographystyle{mn2e}
\bibliography{author2}

\begin{thebibliography}{}

\bibitem[\protect\citeauthoryear{{Allen}, {Schaefer} \& {Scott}}{{Allen}
  et~al.}{2015}]{Allen15}
{Allen} J.~T.,  {Schaefer} A.~L.,    {Scott} N. e.~a.,  2015, \mnras, 451, 2780

\bibitem[\protect\citeauthoryear{{Alpaslan}, {Driver}, {Robotham} \&
  {Obreschkow}}{{Alpaslan} et~al.}{2015}]{Alpaslan15}
{Alpaslan} M.,  {Driver} S.,  {Robotham} A.~S.~G.,    {Obreschkow} D. e.~a.,
  2015, \mnras, 451, 3249

\bibitem[\protect\citeauthoryear{{Alpaslan}, {Grootes}, {Marcum}, {Popescu},
  {Tuffs} \& {Bland-Hawthorn}}{{Alpaslan} et~al.}{2016}]{Alpaslan16}
{Alpaslan} M.,  {Grootes} M.,  {Marcum} P.~M.,  {Popescu} C.,  {Tuffs} R.,
  {Bland-Hawthorn} 2016, \mnras, 457, 2287

\bibitem[\protect\citeauthoryear{{Alpaslan}, {Robotham} \& {Driver}}{{Alpaslan}
  et~al.}{2014}]{Alpaslan14a}
{Alpaslan} M.,  {Robotham} A.~S.~G.,    {Driver} S. e.~a.,  2014, \mnras, 438,
  177

\bibitem[\protect\citeauthoryear{{Arag{\'o}n-Calvo}, {van de Weygaert}, {Jones}
  \& {van der Hulst}}{{Arag{\'o}n-Calvo} et~al.}{2007}]{Aragon07}
{Arag{\'o}n-Calvo} M.~A.,  {van de Weygaert} R.,  {Jones} B.~J.~T.,    {van der
  Hulst} J.~M.,  2007, \apjl, 655, L5

\bibitem[\protect\citeauthoryear{{Aubert}, {Pichon} \& {Colombi}}{{Aubert}
  et~al.}{2004}]{Aubert04}
{Aubert} D.,  {Pichon} C.,    {Colombi} S.,  2004, \mnras, 352, 376

\bibitem[\protect\citeauthoryear{{Baldry}, {Liske}, {Brown} \&
  {Robotham}}{{Baldry} et~al.}{2018}]{Baldry18}
{Baldry} I.~K.,  {Liske} J.,  {Brown} M.~J.~I.,    {Robotham} A.~S.~G. e.~a.,
  2018, \mnras, 474, 3875

\bibitem[\protect\citeauthoryear{{Bett} \& {Frenk}}{{Bett} \&
  {Frenk}}{2012}]{Bett12}
{Bett} P.~E.,  {Frenk} C.~S.,  2012, \mnras, 420, 3324

\bibitem[\protect\citeauthoryear{{Beygu}, {Kreckel}, {van der Hulst} \& et
  al.}{{Beygu} et~al.}{2016}]{Beygu16}
{Beygu} B.,  {Kreckel} K.,  {van der Hulst} J.~M.,    et al. J.,  2016, \mnras,
  458, 394

\bibitem[\protect\citeauthoryear{{Birnboim} \& {Dekel}}{{Birnboim} \&
  {Dekel}}{2003}]{Birnboim03}
{Birnboim} Y.,  {Dekel} A.,  2003, \mnras, 345, 349

\bibitem[\protect\citeauthoryear{{Bland-Hawthorn}, {Bryant} \&
  {Robertson}}{{Bland-Hawthorn} et~al.}{2011}]{BlandHawthorn11}
{Bland-Hawthorn} J.,  {Bryant} J.,    {Robertson} G. e.~a.,  2011, Optics
  Express, 19, 2649

\bibitem[\protect\citeauthoryear{{Bond}, {Kofman} \& {Pogosyan}}{{Bond}
  et~al.}{1996}]{Bond96}
{Bond} J.~R.,  {Kofman} L.,    {Pogosyan} D.,  1996, \nat, 380, 603

\bibitem[\protect\citeauthoryear{{Brainerd}}{{Brainerd}}{2005}]{Brainerd05}
{Brainerd} T.~G.,  2005, \apjl, 628, L101

\bibitem[\protect\citeauthoryear{{Bryant} \& {Bland-Hawthorn}}{{Bryant} \&
  {Bland-Hawthorn}}{2012}]{Bryant12}
{Bryant} J.~J.,  {Bland-Hawthorn} J.,  2012, in Ground-based and Airborne
  Instrumentation for Astronomy IV Vol.~8446 of \procspie, {Square-core bundles
  for astronomical imaging}.
p. 84466K

\bibitem[\protect\citeauthoryear{{Bryant}, {Bland-Hawthorn}, {Fogarty},
  {Lawrence} \& {Croom}}{{Bryant} et~al.}{2014}]{Bryant14}
{Bryant} J.~J.,  {Bland-Hawthorn} J.,  {Fogarty} L.~M.~R.,  {Lawrence} J.~S.,
   {Croom} S.~M.,  2014, \mnras, 438, 869

\bibitem[\protect\citeauthoryear{{Bryant}, {Bland-Hawthorn}, {Lawrence},
  {Croom}, {Brown} \& et al.}{{Bryant} et~al.}{2016}]{Bryant16}
{Bryant} J.~J.,  {Bland-Hawthorn} J.,  {Lawrence} J.,  {Croom} S.,  {Brown} D.,
     et al. V.,  2016, in Ground-based and Airborne Instrumentation for
  Astronomy VI Vol.~9908 of \procspie, {Hector: a new massively multiplexed IFU
  instrument for the Anglo-Australian Telescope}.
p. 99081F

\bibitem[\protect\citeauthoryear{{Bryant}, {O'Byrne}, {Bland-Hawthorn} \&
  {Leon-Saval}}{{Bryant} et~al.}{2011}]{Bryant11}
{Bryant} J.~J.,  {O'Byrne} J.~W.,  {Bland-Hawthorn} J.,    {Leon-Saval} S.~G.,
  2011, \mnras, 415, 2173

\bibitem[\protect\citeauthoryear{{Bryant}, {Owers}, {Robotham} \&
  {Croom}}{{Bryant} et~al.}{2015}]{Bryant15}
{Bryant} J.~J.,  {Owers} M.~S.,  {Robotham} A.~S.~G.,    {Croom} S.~M. e.~a.,
  2015, \mnras, 447, 2857

\bibitem[\protect\citeauthoryear{{Bundy}, {Bershady}, {Law} \& {Yan}}{{Bundy}
  et~al.}{2015}]{Bundy15}
{Bundy} K.,  {Bershady} M.~A.,  {Law} D.~R.,    {Yan} R. e.~a.,  2015, \apj,
  798, 7

\bibitem[\protect\citeauthoryear{{Ca{\~n}as}, {Elahi}, {Welker}, {del P Lagos},
  {Power}, {Dubois} \& {Pichon}}{{Ca{\~n}as} et~al.}{2019}]{Canas18}
{Ca{\~n}as} R.,  {Elahi} P.~J.,  {Welker} C.,  {del P Lagos} C.,  {Power} C.,
  {Dubois} Y.,    {Pichon} C.,  2019, \mnras, 482, 2039

\bibitem[\protect\citeauthoryear{{Cappellari}}{{Cappellari}}{2002}]{Cappellari02}
{Cappellari} M.,  2002, \mnras, 333, 400

\bibitem[\protect\citeauthoryear{{Cappellari}, {Emsellem}, {Bacon}, {Bureau} \&
  {Davies}}{{Cappellari} et~al.}{2007}]{Cappellari07}
{Cappellari} M.,  {Emsellem} E.,  {Bacon} R.,  {Bureau} M.,    {Davies} R.~L.
  e.~a.,  2007, \mnras, 379, 418

\bibitem[\protect\citeauthoryear{{Cappellari}, {van den Bosch} \&
  {Verolme}}{{Cappellari} et~al.}{2004}]{Cappellari04}
{Cappellari} M.,  {van den Bosch} R.~C.~E.,    {Verolme} E.~K. e.~a.,  2004,
  Coevolution of Black Holes and Galaxies, p.~5

\bibitem[\protect\citeauthoryear{{Cautun}, {Bose} \& {Frenk}}{{Cautun}
  et~al.}{2015}]{Cautun15}
{Cautun} M.,  {Bose} S.,    {Frenk} C.~S. e.~a.,  2015, \mnras, 452, 3838

\bibitem[\protect\citeauthoryear{{Cautun}, {van de Weygaert} \&
  {Jones}}{{Cautun} et~al.}{2013}]{Cautun13}
{Cautun} M.,  {van de Weygaert} R.,    {Jones} B.~J.~T.,  2013, \mnras, 429,
  1286

\bibitem[\protect\citeauthoryear{{Chen}, {Ho}, {Blazek}, {He}, {Mandelbaum},
  {Melchior} \& {Singh}}{{Chen} et~al.}{2019}]{Chen19}
{Chen} Y.-C.,  {Ho} S.,  {Blazek} J.,  {He} S.,  {Mandelbaum} R.,  {Melchior}
  P.,    {Singh} S.,  2019, \mnras

\bibitem[\protect\citeauthoryear{{Codis}, {Jindal}, {Chisari}, {Vibert},
  {Dubois}, {Pichon} \& {Devriendt}}{{Codis} et~al.}{2018}]{Codis18}
{Codis} S.,  {Jindal} A.,  {Chisari} N.~E.,  {Vibert} D.,  {Dubois} Y.,
  {Pichon} C.,    {Devriendt} J.,  2018, \mnras, 481, 4753

\bibitem[\protect\citeauthoryear{{Codis}, {Pichon}, {Devriendt}, {Slyz},
  {Pogosyan}, {Dubois} \& {Sousbie}}{{Codis} et~al.}{2012}]{Codis12}
{Codis} S.,  {Pichon} C.,  {Devriendt} J.,  {Slyz} A.,  {Pogosyan} D.,
  {Dubois} Y.,    {Sousbie} T.,  2012, \mnras, 427, 3320

\bibitem[\protect\citeauthoryear{{Codis}, {Pichon} \& {Pogosyan}}{{Codis}
  et~al.}{2015}]{Codis15}
{Codis} S.,  {Pichon} C.,    {Pogosyan} D.,  2015, \mnras, 452, 3369

\bibitem[\protect\citeauthoryear{{Colless}}{{Colless}}{2003}]{Colless03a}
{Colless} M.,  2003, Mercury, 32, 30

\bibitem[\protect\citeauthoryear{{Croom}, {Lawrence} \&
  {Bland-Hawthorn}}{{Croom} et~al.}{2012}]{Croom12}
{Croom} S.~M.,  {Lawrence} J.~S.,    {Bland-Hawthorn} J. e.~a.,  2012, \mnras,
  421, 872

\bibitem[\protect\citeauthoryear{{Danovich}, {Dekel}, {Hahn} \&
  {Teyssier}}{{Danovich} et~al.}{2012}]{Danovich12}
{Danovich} M.,  {Dekel} A.,  {Hahn} O.,    {Teyssier} R.,  2012, \mnras, 422,
  1732

\bibitem[\protect\citeauthoryear{{de Lapparent}, {Geller} \& {Huchra}}{{de
  Lapparent} et~al.}{1986}]{Delapparent86}
{de Lapparent} V.,  {Geller} M.~J.,    {Huchra} J.~P.,  1986, \apjl, 302, L1

\bibitem[\protect\citeauthoryear{{Dekel}, {Birnboim} \& {Engel}}{{Dekel}
  et~al.}{2009}]{Dekel09}
{Dekel} A.,  {Birnboim} Y.,    {Engel} G. e.~a.,  2009, \nat, 457, 451

\bibitem[\protect\citeauthoryear{{Doroshkevich}, {Tucker}, {Allam} \&
  {Way}}{{Doroshkevich} et~al.}{2004}]{Doroshkevich04}
{Doroshkevich} A.,  {Tucker} D.~L.,  {Allam} S.,    {Way} M.~J.,  2004, \aap,
  418, 7

\bibitem[\protect\citeauthoryear{{Driver}, {Hill}, {Kelvin} \&
  {Robotham}}{{Driver} et~al.}{2011}]{Driver11}
{Driver} S.~P.,  {Hill} D.~T.,  {Kelvin} L.~S.,    {Robotham} A.~S.~G. e.~a.,
  2011, \mnras, 413, 971

\bibitem[\protect\citeauthoryear{{Dubois}, {Pichon} \& {Welker}}{{Dubois}
  et~al.}{2014}]{Dubois14}
{Dubois} Y.,  {Pichon} C.,    {Welker} C. e.~a.,  2014, \mnras, 444, 1453

\bibitem[\protect\citeauthoryear{{Eardley}, {Peacock}, {McNaught-Roberts} \&
  {Heymans}}{{Eardley} et~al.}{2015}]{Eardley15}
{Eardley} E.,  {Peacock} J.~A.,  {McNaught-Roberts} T.,    {Heymans} C. e.~a.,
  2015, \mnras, 448, 3665

\bibitem[\protect\citeauthoryear{{Elahi}, {Welker}, {Power} \& {Lagos}}{{Elahi}
  et~al.}{2018}]{Elahi18}
{Elahi} P.~J.,  {Welker} C.,  {Power} C.,    {Lagos} e.~a.,  2018, \mnras, 475,
  5338

\bibitem[\protect\citeauthoryear{{Emsellem}, {Cappellari}, {Krajnovi{\'c}},
  {van de Ven} \& {Bacon}}{{Emsellem} et~al.}{2007}]{Ensellem07}
{Emsellem} E.,  {Cappellari} M.,  {Krajnovi{\'c}} D.,  {van de Ven} G.,
  {Bacon} R. e.~a.,  2007, \mnras, 379, 401

\bibitem[\protect\citeauthoryear{{Emsellem}, {Monnet} \& {Bacon}}{{Emsellem}
  et~al.}{1994}]{Emsellem94}
{Emsellem} E.,  {Monnet} G.,    {Bacon} R.,  1994, \aap, 285, 723

\bibitem[\protect\citeauthoryear{{Green}, {Croom}, {Scott}, {Cortese},
  {Medling}, {D'Eugenio}, {Bryant} \& {Bland-Hawthorn}}{{Green}
  et~al.}{2018}]{Green18}
{Green} A.~W.,  {Croom} S.~M.,  {Scott} N.,  {Cortese} L.,  {Medling} A.~M.,
  {D'Eugenio} F.,  {Bryant} J.~J.,    {Bland-Hawthorn} J. e.~a.,  2018, \mnras,
  475, 716

\bibitem[\protect\citeauthoryear{{Guo}, {Tempel} \& {Libeskind}}{{Guo}
  et~al.}{2015}]{Guo15}
{Guo} Q.,  {Tempel} E.,    {Libeskind} N.~I.,  2015, \apj, 800, 112

\bibitem[\protect\citeauthoryear{{Haardt} \& {Madau}}{{Haardt} \&
  {Madau}}{1996}]{HaardtMadau96}
{Haardt} F.,  {Madau} P.,  1996, \apj, 461, 20

\bibitem[\protect\citeauthoryear{{Hahn}, {Carollo}, {Porciani} \&
  {Dekel}}{{Hahn} et~al.}{2007}]{Hahn07}
{Hahn} O.,  {Carollo} C.~M.,  {Porciani} C.,    {Dekel} A.,  2007, \mnras, 381,
  41

\bibitem[\protect\citeauthoryear{Holmberg}{Holmberg}{1969}]{Holmberg69}
Holmberg E.,  1969, ArA, 5, 305

\bibitem[\protect\citeauthoryear{{Huang}, {Mandelbaum}, {Freeman}, {Chen},
  {Rozo} \& {Rykoff}}{{Huang} et~al.}{2017}]{Huang17}
{Huang} H.-J.,  {Mandelbaum} R.,  {Freeman} P.~E.,  {Chen} Y.-C.,  {Rozo} E.,
   {Rykoff} E.,  2017, ArXiv e-prints

\bibitem[\protect\citeauthoryear{{Kaviraj}, {Laigle}, {Kimm}, {Devriendt},
  {Dubois}, {Pichon}, {Slyz}, {Chisari} \& {Peirani}}{{Kaviraj}
  et~al.}{2017}]{Kaviraj17}
{Kaviraj} S.,  {Laigle} C.,  {Kimm} T.,  {Devriendt} J.~E.~G.,  {Dubois} Y.,
  {Pichon} C.,  {Slyz} A.,  {Chisari} E.,    {Peirani} S.,  2017, \mnras, 467,
  4739

\bibitem[\protect\citeauthoryear{{Kleiner}, {Pimbblet}, {Jones}, {Koribalski}
  \& {Serra}}{{Kleiner} et~al.}{2017}]{Kleiner17}
{Kleiner} D.,  {Pimbblet} K.~A.,  {Jones} D.~H.,  {Koribalski} B.~S.,
  {Serra} P.,  2017, \mnras, 466, 4692

\bibitem[\protect\citeauthoryear{{Knebe}, {Gill} \& {Gibson}}{{Knebe}
  et~al.}{2004}]{Knebe04}
{Knebe} A.,  {Gill} S.~P.~D.,    {Gibson} B.~K. e.~a.,  2004, \apj, 603, 7

\bibitem[\protect\citeauthoryear{{Komatsu}, {Smith}, {Dunkley}, {Bennett} \&
  {Gold}}{{Komatsu} et~al.}{2011}]{Komatsu11}
{Komatsu} E.,  {Smith} K.~M.,  {Dunkley} J.,  {Bennett} C.~L.,    {Gold} B.
  e.~a.,  2011, \apjs, 192, 18

\bibitem[\protect\citeauthoryear{{Krajnovi{\'c}}, {Cappellari}, {de Zeeuw} \&
  {Copin}}{{Krajnovi{\'c}} et~al.}{2006}]{Krajnovic06}
{Krajnovi{\'c}} D.,  {Cappellari} M.,  {de Zeeuw} P.~T.,    {Copin} Y.,  2006,
  \mnras, 366, 787

\bibitem[\protect\citeauthoryear{{Kraljic}, {Arnouts}, {Pichon} \&
  {Laigle}}{{Kraljic} et~al.}{2018}]{Kraljic18}
{Kraljic} K.,  {Arnouts} S.,  {Pichon} C.,    {Laigle} C. e.~a.,  2018, \mnras,
  474, 547

\bibitem[\protect\citeauthoryear{{Kraljic}, {Pichon}, {Dubois} \&
  {Codis}}{{Kraljic} et~al.}{2018}]{Kraljic18b}
{Kraljic} K.,  {Pichon} C.,  {Dubois} Y.,    {Codis} S. e.~a.,  2018, ArXiv
  e-prints

\bibitem[\protect\citeauthoryear{{Krolewski}, {Ho} \& {Chen}}{{Krolewski}
  et~al.}{2019}]{Krowleski19}
{Krolewski} A.,  {Ho} S.,    {Chen} Y. e.~a.,  2019, arxiv e-print

\bibitem[\protect\citeauthoryear{{Kuutma}, {Tamm} \& {Tempel}}{{Kuutma}
  et~al.}{2017}]{Kuutma17}
{Kuutma} T.,  {Tamm} A.,    {Tempel} E. e.~a.,  2017, \aap, 600, L6

\bibitem[\protect\citeauthoryear{{Lagos}, {Tobar}, {Robotham} \&
  {Obreschkow}}{{Lagos} et~al.}{2018}]{Lagos18}
{Lagos} C.~d.~P.,  {Tobar} R.~J.,  {Robotham} A.~S.~G.,    {Obreschkow} D.
  e.~a.,  2018, \mnras, 481, 3573

\bibitem[\protect\citeauthoryear{{Laigle}, {Davidzon}, {Ilbert}, {Devriendt},
  {Kashino} \& {Pichon}}{{Laigle} et~al.}{2019}]{Laigle19}
{Laigle} C.,  {Davidzon} I.,  {Ilbert} O.,  {Devriendt} J.,  {Kashino} D.,
  {Pichon} C. e.~a.,  2019, arXiv e-prints

\bibitem[\protect\citeauthoryear{{Laigle}, {Pichon}, {Arnouts} \&
  {McCracken}}{{Laigle} et~al.}{2018}]{Laigle18}
{Laigle} C.,  {Pichon} C.,  {Arnouts} S.,    {McCracken} H.~J. e.~a.,  2018,
  \mnras, 474, 5437

\bibitem[\protect\citeauthoryear{{Laigle}, {Pichon} \& {Codis}}{{Laigle}
  et~al.}{2015}]{Laigle15}
{Laigle} C.,  {Pichon} C.,    {Codis} S. e.~a.,  2015, \mnras, 446, 2744

\bibitem[\protect\citeauthoryear{{Libeskind}, {van de Weygaert} \&
  {Cautun}}{{Libeskind} et~al.}{2018}]{Libeskind18}
{Libeskind} N.~I.,  {van de Weygaert} R.,    {Cautun} M. e.~a.,  2018, \mnras,
  473, 1195

\bibitem[\protect\citeauthoryear{{Malavasi}, {Arnouts}, {Vibert} \& {de la
  Torre}}{{Malavasi} et~al.}{2017}]{Malavasi17}
{Malavasi} N.,  {Arnouts} S.,  {Vibert} D.,    {de la Torre} S. e.~a.,  2017,
  \mnras, 465, 3817

\bibitem[\protect\citeauthoryear{{Nierenberg}, {Auger}, {Treu} \&
  {Marshall}}{{Nierenberg} et~al.}{2012}]{Nierenberg12}
{Nierenberg} A.~M.,  {Auger} M.~W.,  {Treu} T.,    {Marshall} P.~J. e.~a.,
  2012, \apj, 752, 99

\bibitem[\protect\citeauthoryear{{Ocvirk}, {Pichon} \& {Teyssier}}{{Ocvirk}
  et~al.}{2008}]{Ocvirk08}
{Ocvirk} P.,  {Pichon} C.,    {Teyssier} R.,  2008, \mnras, 390, 1326

\bibitem[\protect\citeauthoryear{{Owers}, {Allen}, {Baldry} \&
  {Bryant}}{{Owers} et~al.}{2017}]{Owers17}
{Owers} M.~S.,  {Allen} J.~T.,  {Baldry} I.,    {Bryant} J.~J. e.~a.,  2017,
  \mnras, 468, 1824

\bibitem[\protect\citeauthoryear{{Pahwa}, {Libeskind}, {Tempel}, {Hoffman},
  {Tully}, {Courtois}, {Gottl{\"o}ber}, {Steinmetz} \& {Sorce}}{{Pahwa}
  et~al.}{2016}]{Pahwa16}
{Pahwa} I.,  {Libeskind} N.~I.,  {Tempel} E.,  {Hoffman} Y.,  {Tully} R.~B.,
  {Courtois} H.~M.,  {Gottl{\"o}ber} S.,  {Steinmetz} M.,    {Sorce} J.~G.,
  2016, \mnras, 457, 695

\bibitem[\protect\citeauthoryear{{Paz}, {Stasyszyn} \& {Padilla}}{{Paz}
  et~al.}{2008}]{Paz08}
{Paz} D.~J.,  {Stasyszyn} F.,    {Padilla} N.~D.,  2008, \mnras, 389, 1127

\bibitem[\protect\citeauthoryear{{Peebles}}{{Peebles}}{1980}]{Peebles80}
{Peebles} P.~J.~E.,  1980, {The large-scale structure of the universe}

\bibitem[\protect\citeauthoryear{{Pichon}, {Pogosyan}, {Kimm}, {Slyz},
  {Devriendt} \& {Dubois}}{{Pichon} et~al.}{2011}]{Pichon11}
{Pichon} C.,  {Pogosyan} D.,  {Kimm} T.,  {Slyz} A.,  {Devriendt} J.,
  {Dubois} Y.,  2011, \mnras, 418, 2493

\bibitem[\protect\citeauthoryear{{Porciani}, {Dekel} \& {Hoffman}}{{Porciani}
  et~al.}{2002}]{Porciani02}
{Porciani} C.,  {Dekel} A.,    {Hoffman} Y.,  2002, \mnras, 332, 325

\bibitem[\protect\citeauthoryear{{Poulton}, {Robotham}, {Power} \&
  {Elahi}}{{Poulton} et~al.}{2018}]{Poulton18}
{Poulton} R.~J.~J.,  {Robotham} A.~S.~G.,  {Power} C.,    {Elahi} P.~J.,  2018,
  ArXiv e-prints

\bibitem[\protect\citeauthoryear{{Rojas}, {Vogeley}, {Hoyle} \&
  {Brinkmann}}{{Rojas} et~al.}{2004}]{Rojas04}
{Rojas} R.~R.,  {Vogeley} M.~S.,  {Hoyle} F.,    {Brinkmann} J.,  2004, \apj,
  617, 50

\bibitem[\protect\citeauthoryear{{Sales} \& {Lambas}}{{Sales} \&
  {Lambas}}{2009}]{Sales09}
{Sales} L.,  {Lambas} D.~G.,  2009, \mnras, 395, 1184

\bibitem[\protect\citeauthoryear{{S{\'a}nchez-Bl{\'a}zquez}, {Gorgas},
  {Cardiel} \& {Gonz{\'a}lez}}{{S{\'a}nchez-Bl{\'a}zquez}
  et~al.}{2006}]{SanchezBlazquez06}
{S{\'a}nchez-Bl{\'a}zquez} P.,  {Gorgas} J.,  {Cardiel} N.,    {Gonz{\'a}lez}
  J.~J.,  2006, \aap, 457, 787

\bibitem[\protect\citeauthoryear{{Saunders}, {Bridges} \&
  {Gillingham}}{{Saunders} et~al.}{2004}]{Saunders04}
{Saunders} W.,  {Bridges} T.,    {Gillingham} P. e.~a.,  2004, in {Moorwood}
  A.~F.~M.,  {Iye} M.,  eds, Ground-based Instrumentation for Astronomy
  Vol.~5492 of \procspie, {AAOmega: a scientific and optical overview}.
pp 389--400

\bibitem[\protect\citeauthoryear{{Scott}, {Cappellari}, {Davies}, {Verdoes
  Kleijn}, {Bois} \& {Alatalo}}{{Scott} et~al.}{2013}]{Scott13}
{Scott} N.,  {Cappellari} M.,  {Davies} R.~L.,  {Verdoes Kleijn} G.,  {Bois}
  M.,    {Alatalo} K. e.~a.,  2013, \mnras, 432, 1894

\bibitem[\protect\citeauthoryear{{Scott}, {van de Sande}, {Croom}, {Groves} \&
  {Owers}}{{Scott} et~al.}{2018}]{Scott18}
{Scott} N.,  {van de Sande} J.,  {Croom} S.~M.,  {Groves} B.,    {Owers} M.~S.
  e.~a.,  2018, \mnras, 481, 2299

\bibitem[\protect\citeauthoryear{{Shandarin} \& {Zeldovich}}{{Shandarin} \&
  {Zeldovich}}{1989}]{Shandarin89}
{Shandarin} S.~F.,  {Zeldovich} Y.~B.,  1989, Reviews of Modern Physics, 61,
  185

\bibitem[\protect\citeauthoryear{{Shanks}, {Belokurov}, {Chehade} \&
  {Croom}}{{Shanks} et~al.}{2013}]{Shanks13}
{Shanks} T.,  {Belokurov} V.,  {Chehade} B.,    {Croom} S.~M. e.~a.,  2013, The
  Messenger, 154, 38

\bibitem[\protect\citeauthoryear{{Sharp}, {Allen} \& {Fogarty}}{{Sharp}
  et~al.}{2015}]{Sharp15}
{Sharp} R.,  {Allen} J.~T.,    {Fogarty} L.~M.~R. e.~a.,  2015, \mnras, 446,
  1551

\bibitem[\protect\citeauthoryear{{Sharp}, {Saunders} \& {Smith}}{{Sharp}
  et~al.}{2006}]{Sharp06}
{Sharp} R.,  {Saunders} W.,    {Smith} G. e.~a.,  2006, in Society of
  Photo-Optical Instrumentation Engineers (SPIE) Conference Series Vol.~6269 of
  \procspie, {Performance of AAOmega: the AAT multi-purpose fiber-fed
  spectrograph}.
p. 62690G

\bibitem[\protect\citeauthoryear{{Shivshankar}, {Pranav}, {Natarajan}, {van de
  Weygaert}, {Bos} \& {Rieder}}{{Shivshankar} et~al.}{2015}]{Felix15}
{Shivshankar} N.,  {Pranav} P.,  {Natarajan} V.,  {van de Weygaert} R.,  {Bos}
  E.~P.,    {Rieder} S.,  2015, arXiv e-prints

\bibitem[\protect\citeauthoryear{{Smith}, {Saunders} \& {Bridges}}{{Smith}
  et~al.}{2004}]{Smith04}
{Smith} G.~A.,  {Saunders} W.,    {Bridges} T. e.~a.,  2004, in {Moorwood}
  A.~F.~M.,  {Iye} M.,  eds, Ground-based Instrumentation for Astronomy
  Vol.~5492 of \procspie, {AAOmega: a multipurpose fiber-fed spectrograph for
  the AAT}.
pp 410--420

\bibitem[\protect\citeauthoryear{{Smith}, {Duc}, {Bournaud} \& {Yi}}{{Smith}
  et~al.}{2015}]{smithetal15}
{Smith} R.,  {Duc} P.-A.,  {Bournaud} F.,    {Yi} S.~K.,  2015, ArXiv e-prints

\bibitem[\protect\citeauthoryear{{Sousbie}}{{Sousbie}}{2011}]{Sousbie11}
{Sousbie} T.,  2011, \mnras, 414, 350

\bibitem[\protect\citeauthoryear{{Tempel} \& {Libeskind}}{{Tempel} \&
  {Libeskind}}{2013}]{Tempel13b}
{Tempel} E.,  {Libeskind} N.~I.,  2013, \apjl, 775, L42

\bibitem[\protect\citeauthoryear{{Tempel}, {Stoica} \& {Saar}}{{Tempel}
  et~al.}{2013}]{Tempel13a}
{Tempel} E.,  {Stoica} R.~S.,    {Saar} E.,  2013, \mnras, 428, 1827

\bibitem[\protect\citeauthoryear{{Teyssier}}{{Teyssier}}{2002}]{Teyssier02}
{Teyssier} R.,  2002, \aap, 385, 337

\bibitem[\protect\citeauthoryear{{Tweed}, {Devriendt}, {Blaizot}, {Colombi} \&
  {Slyz}}{{Tweed} et~al.}{2009}]{Tweed09}
{Tweed} D.,  {Devriendt} J.,  {Blaizot} J.,  {Colombi} S.,    {Slyz} A.,  2009,
  \aap, 506, 647

\bibitem[\protect\citeauthoryear{{van de Sande}, {Bland-Hawthorn} \&
  {Fogarty}}{{van de Sande} et~al.}{2017a}]{VdS17b}
{van de Sande} J.,  {Bland-Hawthorn} J.,    {Fogarty} L.~M.~R. e.~a.,  2017a,
  \apj, 835, 104

\bibitem[\protect\citeauthoryear{{van de Sande}, {Bland-Hawthorn} \&
  {Fogarty}}{{van de Sande} et~al.}{2017b}]{VdS17a}
{van de Sande} J.,  {Bland-Hawthorn} J.,    {Fogarty} L.~M.~R. e.~a.,  2017b,
  VizieR Online Data Catalog, 183

\bibitem[\protect\citeauthoryear{{van de Sande}, {Lagos} \& {Welker}}{{van de
  Sande} et~al.}{2018}]{VdS18}
{van de Sande} J.,  {Lagos} C.~D.~P.,    {Welker} C. e.~a.,  2018, ArXiv
  e-prints

\bibitem[\protect\citeauthoryear{{Wang}, {Jing}, {Mao} \& {Kang}}{{Wang}
  et~al.}{2005}]{Wang05}
{Wang} H.~Y.,  {Jing} Y.~P.,  {Mao} S.,    {Kang} X.,  2005, \mnras, 364, 424

\bibitem[\protect\citeauthoryear{{Wang}, {Park}, {Hwang} \& {Chen}}{{Wang}
  et~al.}{2010}]{Wang10}
{Wang} Y.,  {Park} C.,  {Hwang} H.~S.,    {Chen} X.,  2010, \apj, 718, 762

\bibitem[\protect\citeauthoryear{{Welker}, {Devriendt}, {Dubois}, {Pichon} \&
  {Peirani}}{{Welker} et~al.}{2014}]{Welker14}
{Welker} C.,  {Devriendt} J.,  {Dubois} Y.,  {Pichon} C.,    {Peirani} S.,
  2014, \mnras, 445, L46

\bibitem[\protect\citeauthoryear{{Welker}, {Dubois}, {Pichon}, {Devriendt} \&
  {Chisari}}{{Welker} et~al.}{2015}]{Welker17}
{Welker} C.,  {Dubois} Y.,  {Pichon} C.,  {Devriendt} J.,    {Chisari} E.~N.,
  2015, ArXiv e-prints

\bibitem[\protect\citeauthoryear{{Welker}, {Dubois}, {Pichon}, {Devriendt} \&
  {Chisari}}{{Welker} et~al.}{2018}]{Welker18}
{Welker} C.,  {Dubois} Y.,  {Pichon} C.,  {Devriendt} J.,    {Chisari} N.~E.,
  2018, \aap, 613, A4

\bibitem[\protect\citeauthoryear{{Yang}, {van den Bosch}, {Mo}, {Mao} \&
  {Kang}}{{Yang} et~al.}{2006}]{Yang06}
{Yang} X.,  {van den Bosch} F.~C.,  {Mo} H.~J.,  {Mao} S.,    {Kang} X.,  2006,
  \mnras, 369, 1293

\bibitem[\protect\citeauthoryear{{York}, {Adelman}, {Anderson} \& {SDSS
  Collaboration}}{{York} et~al.}{2000}]{York00}
{York} D.~G.,  {Adelman} J.,  {Anderson} J. e.~a.,    {SDSS Collaboration}
  2000, \aj, 120, 1579

\bibitem[\protect\citeauthoryear{{Zaritsky}, {Smith}, {Frenk} \&
  {White}}{{Zaritsky} et~al.}{1997}]{Zaritsky97}
{Zaritsky} D.,  {Smith} R.,  {Frenk} C.~S.,    {White} S.~D.~M.,  1997, \apjl,
  478, L53

\bibitem[\protect\citeauthoryear{{Zel'dovich}}{{Zel'dovich}}{1970}]{Zeldo70}
{Zel'dovich} Y.~B.,  1970, \aap, 5, 84

\bibitem[\protect\citeauthoryear{{Zentner}, {Kravtsov}, {Gnedin} \&
  {Klypin}}{{Zentner} et~al.}{2005}]{Zentner05}
{Zentner} A.~R.,  {Kravtsov} A.~V.,  {Gnedin} O.~Y.,    {Klypin} A.~A.,  2005,
  \apj, 629, 219

\end{thebibliography}


\appendix

\section{Stellar mass distribution}
\label{section:mass-distr}

\begin{figure}
\center \includegraphics[width=0.9\columnwidth]{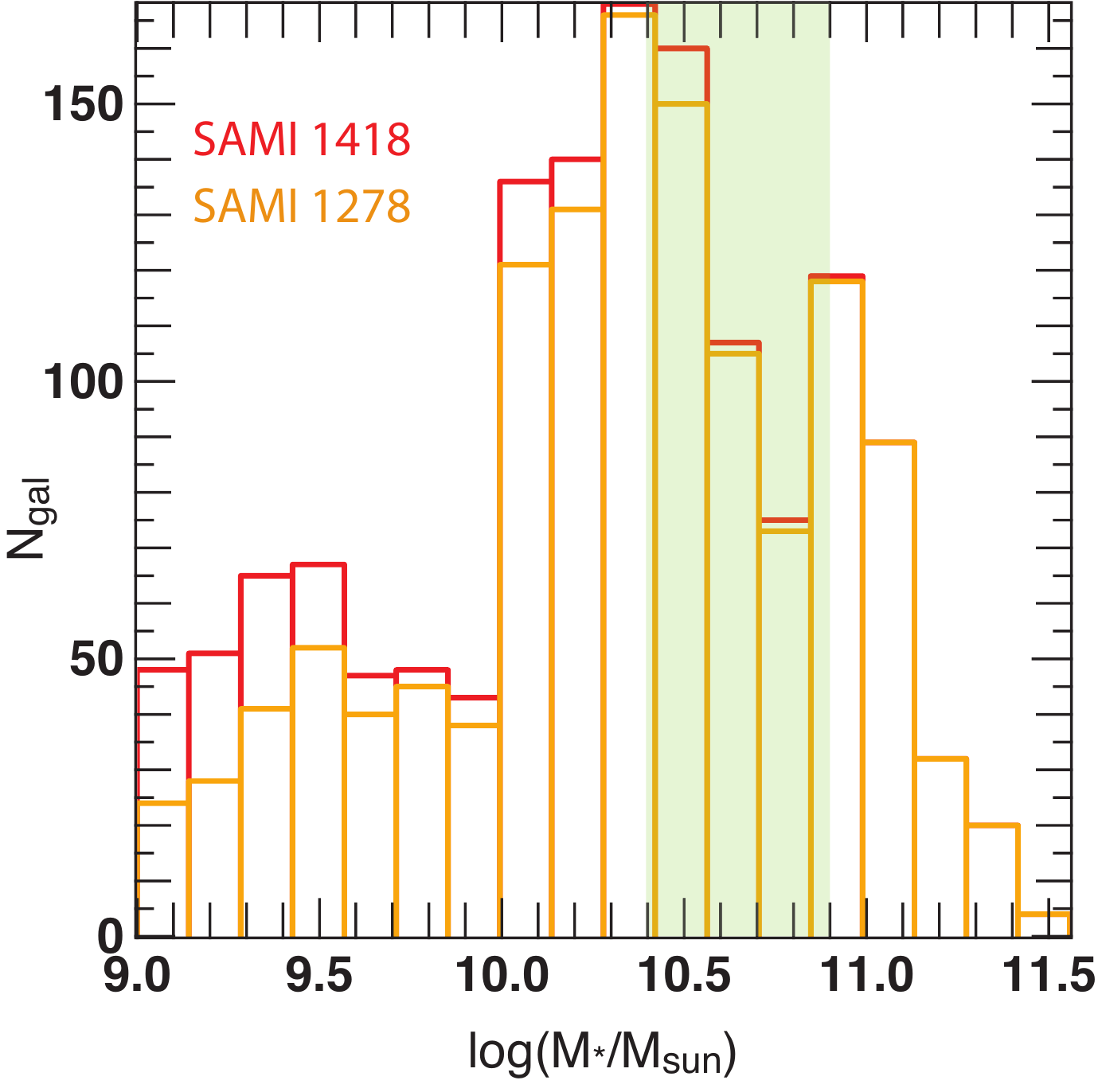}
  \caption{Histogram of the stellar mass distribution in the SAMI sample used in this analysis. The red histogram shows the distribution for the 1418 galaxies of the main sample, the orange is limited to the 1278 galaxies with an effective radius and maximum measurable radius bigger than the seeing HWHM. The green shaded area shows the range in which the spin alignment transition mass is found.}
\label{fig:mdistr}
\end{figure}

Fig.~\ref{fig:mdistr} displays the histogram of the stellar mass distribution in the SAMI sample used in our analysis. The red histogram shows the distribution for the 1418 galaxies of the main sample, the orange is limited to the 1278 galaxies with an effective radius and maximum measurable radius bigger than the seeing HWHM. The green shaded area shows the range in which the spin alignment transition mass is found.

It is apparent that the transition mass range is well sampled in the SAMI sample. As expected, focusing on the most resolved galaxies with effective radius and maximum measurable radius bigger than the seeing HWHM essentially trims out the low-mass range ($M_{*}<10^{10.3}\,M_{\odot}$) of its smallest galaxies (typically $R_{\rm e}< 2$ kpc).

\section{Evolution of stellar mass with distance to filament in Horizon-AGN.}
\label{section:SAMI2}

\begin{figure}
\center \includegraphics[width=1\columnwidth]{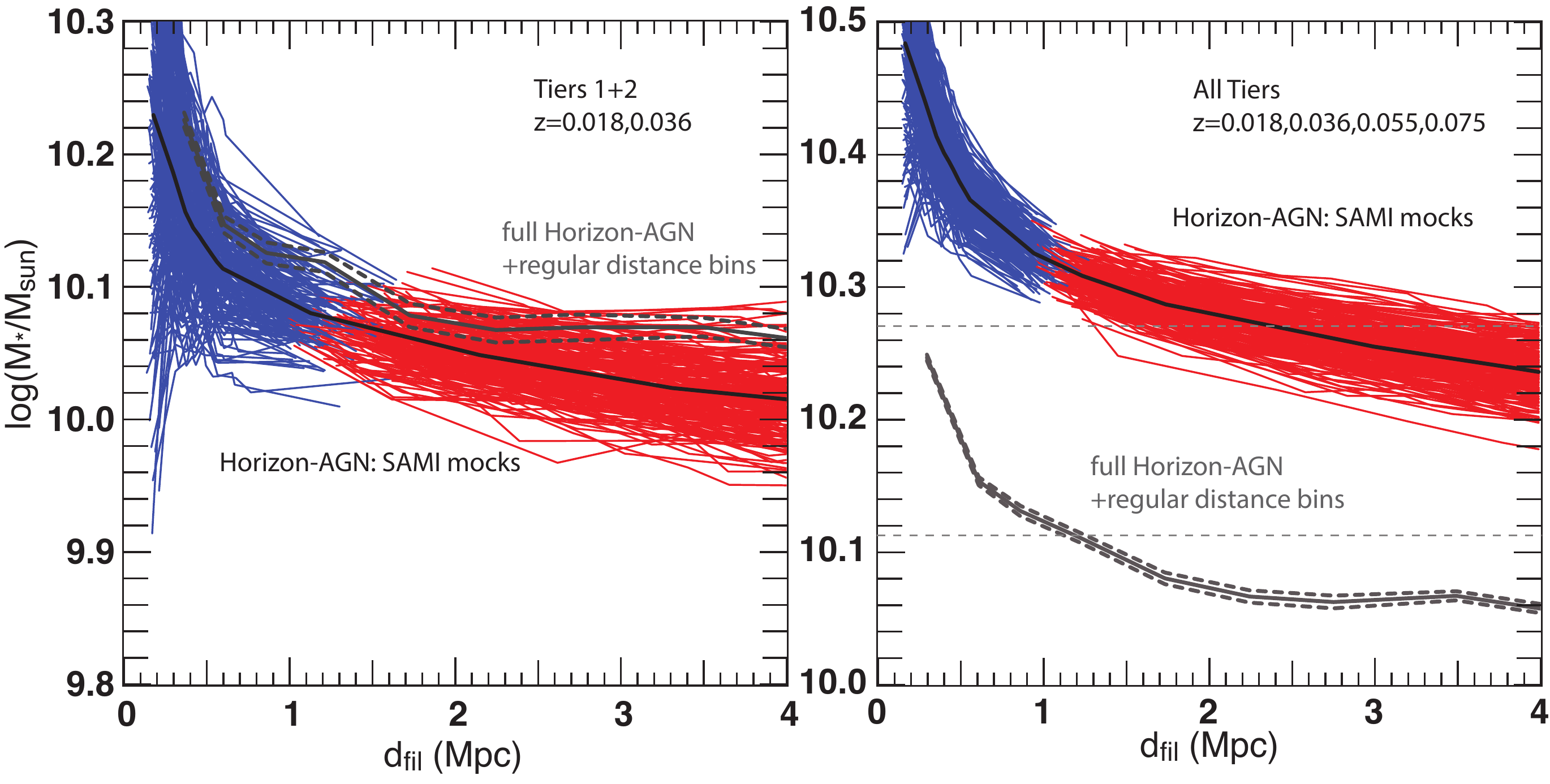}
  \caption{Evolution of average stellar mass with distance to filament for 200 SAMI mocks synthesized from Horizon-AGN (red and blue curves) using shifting distance bins and for the full Horizon-AGN distribution (grey curves) using regular distance bins.  Although average stellar masses vary in different samples due to variations of the corresponding mass function, the relative evolution over distance is preserved and similar to that found in SAMI.}
\label{fig:mhagn}
\end{figure}

Fig.~\ref{fig:mhagn} displays the evolution of the average stellar mass with distance to filaments in the Horizon-AGN simulation using the gas filaments. The solid grey line shows the evolution obtained for the full galaxy population in Horizon-AGN using regular contiguous distance bins. Dashed lines indicate the typical error on the mean. 

Blue and red lines show the same evolution obtained for 200 SAMI mock samples, i.e. sub-samples of Horizon-AGN with statistics matched to SAMI and adapted stellar mass distribution (with the caveat that only galaxies with $M_{*}>10^{9.5}\,M_{\odot}$ can be used in the simulation). We also use for each mock sample the same shifting bin technic  used for SAMI. The black solid line shows the average evolution for all the samples. 

In both cases, with the exceptions of offsets due to variations in the mass function, the evolution is extremely similar for the SAMI mocks and for the full Horizon-AGN population: the average stellar mass is maximal at the centre of the filaments and decreases outwards. This is fully comparable to the evolution found in the SAMI sample, confirming the good representativity of this observed sample in terms of distribution around the cosmic web.

\section{Evolution of alignments with stellar mass in Horizon-AGN. }
\label{section:evolm}

\begin{figure*}
\center \includegraphics[width=2\columnwidth]{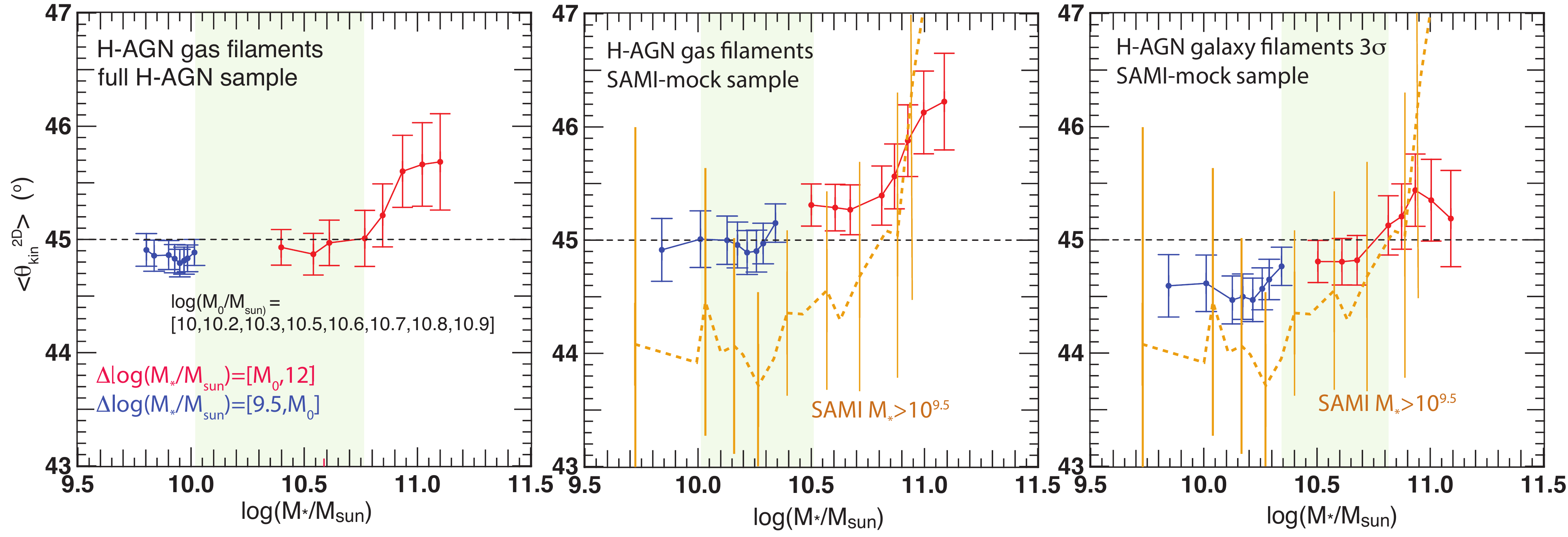}
  \caption{Evolution of $\langle \theta_{\rm kin}^{\rm 2D} \rangle$ (in each bin) as a function of the median mass in the bin $\delta M_{*}$. Low-mass bins are plotted in blue while high-mass bins appear in red.$\langle \theta_{\rm kin}^{\rm 2D} \rangle$ increases steadily with median stellar mass. The horizontal dashed line shows the expectation for uniformly distributed angles. The orange shaded area shows the range in which the transition mass can be bracketed. {\it Left panel:}: Full SAMI population. {\it Intermediate panel:}: Full Horizon-AGN population with gas filaments. {\it Right panel:}: SAMI mock sample population with galaxy filaments. }
\label{fig:hagn-m}
\end{figure*}

To better constrain the transition mass in Horizon-AGN, Fig.~\ref{fig:hagn-m} reproduces an analysis performed on the SAMI sample: it displays the average dependance of $\langle \theta_{\rm kin}^{\rm 2D} \rangle$ on stellar mass for the three Horizon-AGN 
sample. 
Mass bins are defined similarly to what was done in SAMI, with adapted edge values:
\begin{itemize}
\item ``Blue" bins: each bin contains all galaxies with a stellar mass $10^{9.5}\,M_{\odot}<M_{*}<10^{M_{\rm i}}\,M_{\odot}$ with $M_{\rm i}$ taking values 10.1, 10.3, 10.5, 10.6, 10.7, 10.8, 10.9, 11 and 12.
\item ``Red" bins: each bin contains all galaxies with a stellar mass $10^{M_{\rm i}}\,M_{\odot}<M_{*}<10^{12}\,M_{\odot}$ with $M_{\rm i}$ taking values 9.5, 10, 10.2, 10.3, 10.5, 10.6, 10.7, 10.8, 10.9, 11.
\end{itemize}
For comparison, we overlay SAMI results restricted to galaxies with $ M_{*}>10^{9.5}\,M_{\odot}$ in brown orange. Green shaded areas indicate the range within which the transition mass can be bracketed for each population. This mass is overall found to fall within the range $10^{10}\,M_{\odot}<M_{\rm thresh}<10^{10.8}\,M_{\odot}$, consistently with values found in SAMI. The use of gas filaments seems to bias the transition mass towards the lower part of the range.

We find an evolution that is qualitatively similar to that observed in SAMI: the average spin-filament angle progressively transition from values lower than $45^{o}$ to values higher than $45^{o}$ as the median stellar mass increases from $10^{10}$ to $10^{11}\,M_{\odot}$.

\section{Effect of spatial resolution in SAMI.}
\label{section:spatial}

Following Fig.~\ref{fig:thetahagn}, Fig.~\ref{fig:sami1289} displays the average angle in the low-mass subsample $\langle \theta_{\rm low} \rangle$ versus the average angle in the high-mass subsample $\langle \theta_{\rm high} \rangle$ calculated for the Horizon-AGN SAMI-mock population and galaxy filaments. Straight black dashed lines show the expectations in each sample for uniformly distributed angles ($45^{\rm o}$). The black open circle and orange contours show the results for the Horizon-AGN sample considered. We varied the mass threshold as we did for SAMI and present results obtained for the mass threshold that maximises signal-to-noise while maintaining each individual signal above the marginalized $1-\sigma$ threshold: $M_{\rm thresh}=10^{10.7}\, M_{\odot}$. Red-orange contours are computed using bootstrap. Darker to lighter shades indicate the regions in which $50 \%$, $68 \%$  and $90 \%$  of such signals lie.  

In addition, we overlay various SAMI results obtained for the restricted sample of 1278 galaxies with an effective radius and maximum measurable radius bigger than the seeing HWHM. Corresponding 1-$\sigma$ contours are overlaid in pink, green and purple for comparison.

\begin{figure}
\center \includegraphics[width=0.9\columnwidth]{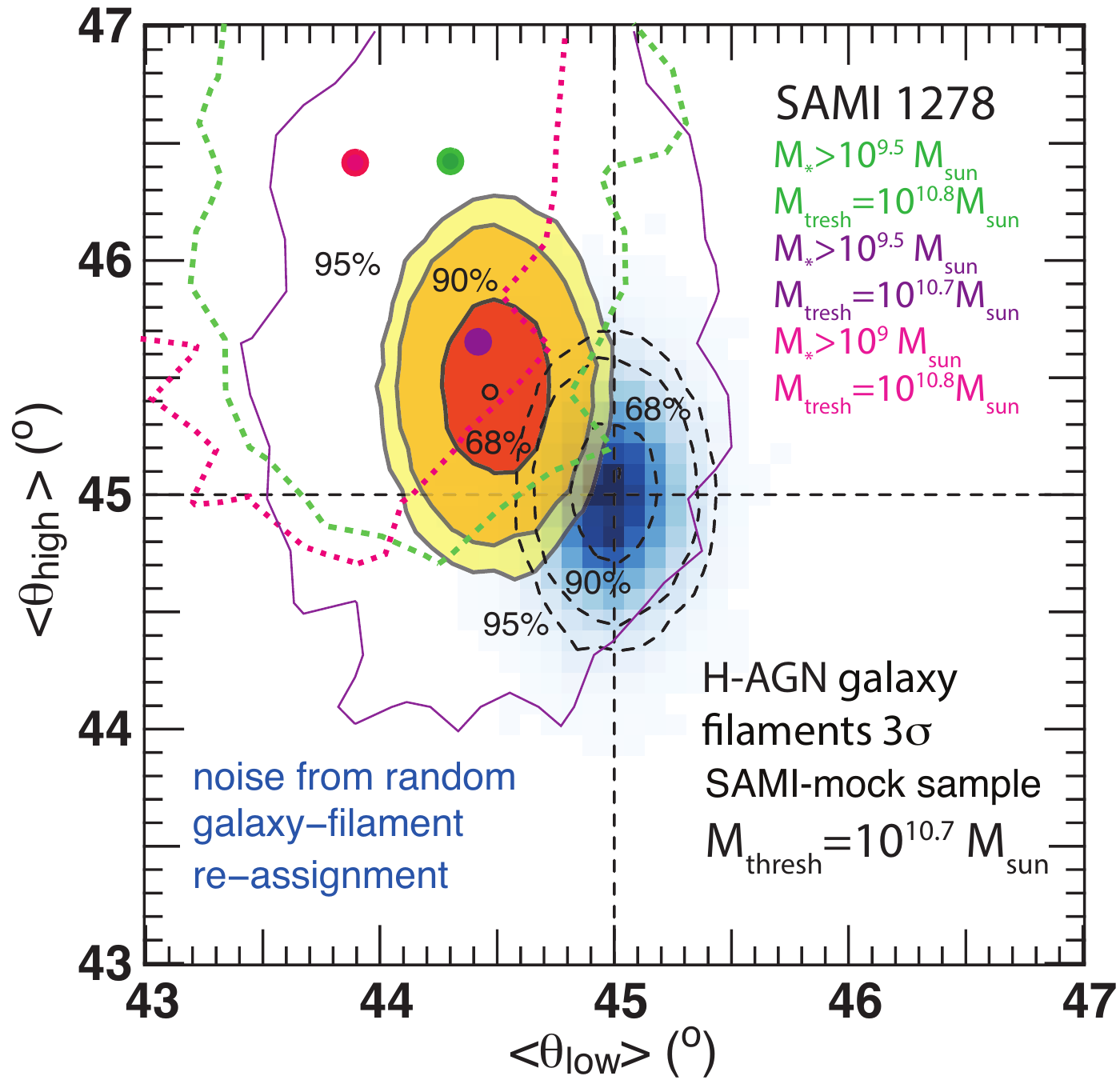}
  \caption{Average spin-filament angle for galaxies with $M_{*}< M_{\rm thresh}$ ("low") versus average angle for galaxies with $M_{*}> M_{\rm thresh}$ ("high") in the Horizon-AGN simulation for the SAMI-mock population with $3\sigma$ galaxy filaments (orange contours). Shades of blue and dashed contours indicate the distribution of values for the expected level of noise. SAMI 1278 signals with specific sub-samples and mass thresholds are overlaid with 1-$\sigma$ contours in pink, green and purple }
\label{fig:sami1289}
\end{figure}

Interestingly, trimming the full SAMI sample of its smallest galaxies tends to reduce the low-mass alignment signal and bring it to values consistent with Horizon-AGN. This confirms the importance of the smallest, lowest mass galaxies in the sample to recover a strong alignment signal. Since the smallest galaxies in Horizon-AGN have a poorly resolved angular momentum as an effect of limited spatial resolution -- $1$ kpc --, hence cannot contribute to a coherent alignment trend, this explains the decreased alignment signal in Horizon-AGN compared to SAMI galaxies.

\section{Effects of refinement level of the filamentary network.}
\label{section:refine}

In this section, we investigate the effect of refining the re-construction of the GAMA cosmic web to reveal less robust, smaller scale filaments on the alignment signals.

\begin{figure}
\center \includegraphics[width=0.95\columnwidth]{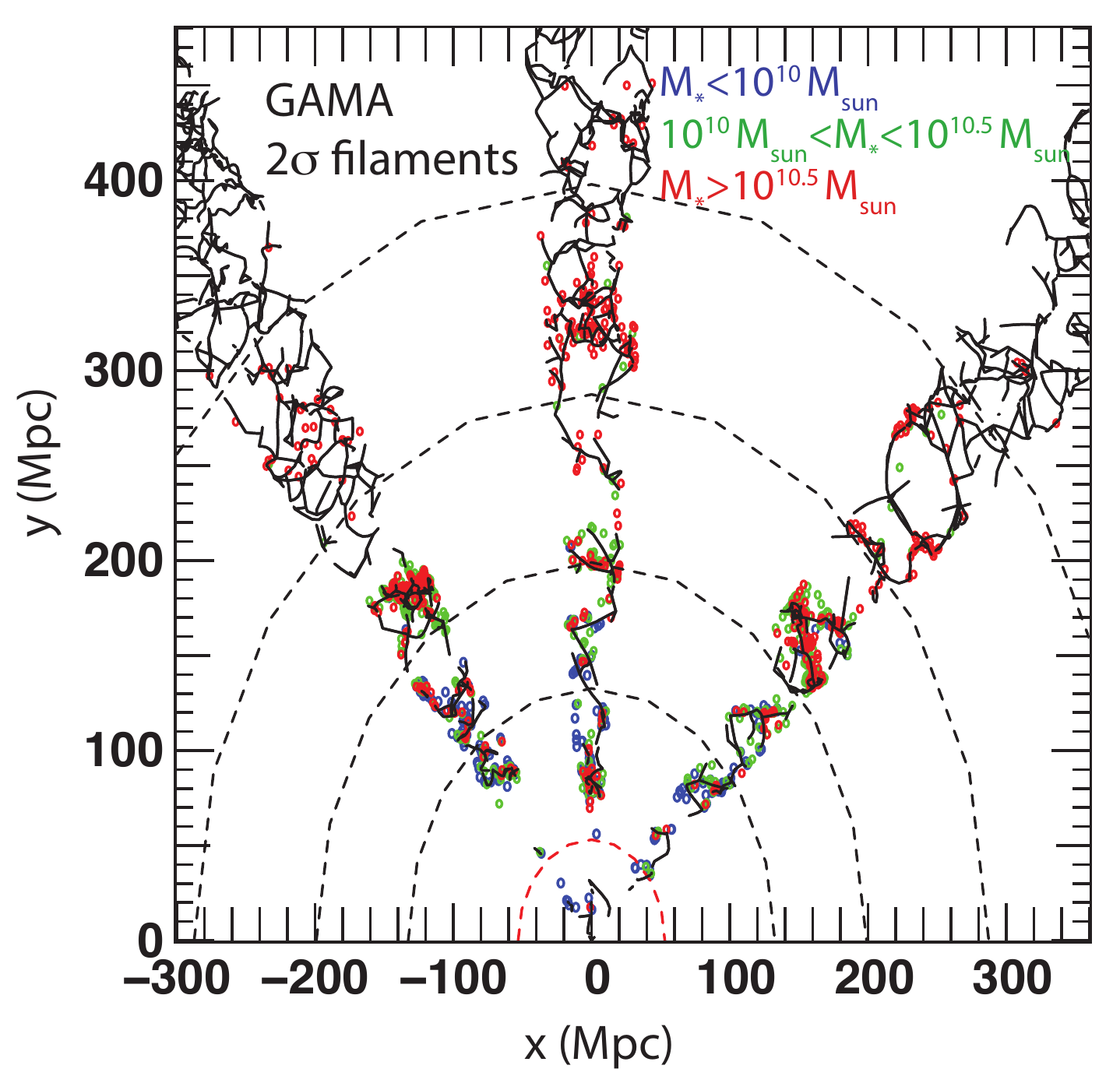}
  \caption{Projected reconstructed network of cosmic filaments across the three GAMA fields that host SAMI galaxies (solid black lines) for a persistence cute of $2\sigma$. SAMI galaxies with $M_{*}>10^{10.5}\, M_{\odot}$ are indicated as red circles, those with $M_{*}<10^{10}\, M_{\odot}$ as blue circles and intermediate range galaxies appear in green. Dashed hemicircles indicate the redshift tiers of the SAMI survey.}
\label{fig:cw2s}
\end{figure}

Fig.~\ref{fig:cw2s} shows a real-space map of the projected reconstructed network of cosmic filaments across the three GAMA fields that host SAMI galaxies (solid black lines) for a persistence cut of $2\sigma$ instead of $3\sigma$. As in the $3\sigma$ map, SAMI galaxies with $M_{*}>10^{10.5}\, M_{\odot}$ are indicated as red circles, those with $M_{*}<10^{10}\, M_{\odot}$ as blue circles and others as green circles. Dashed hemicircles indicate the redshift tiers of the SAMI survey.
\begin{figure*}
\center \includegraphics[width=1.9\columnwidth]{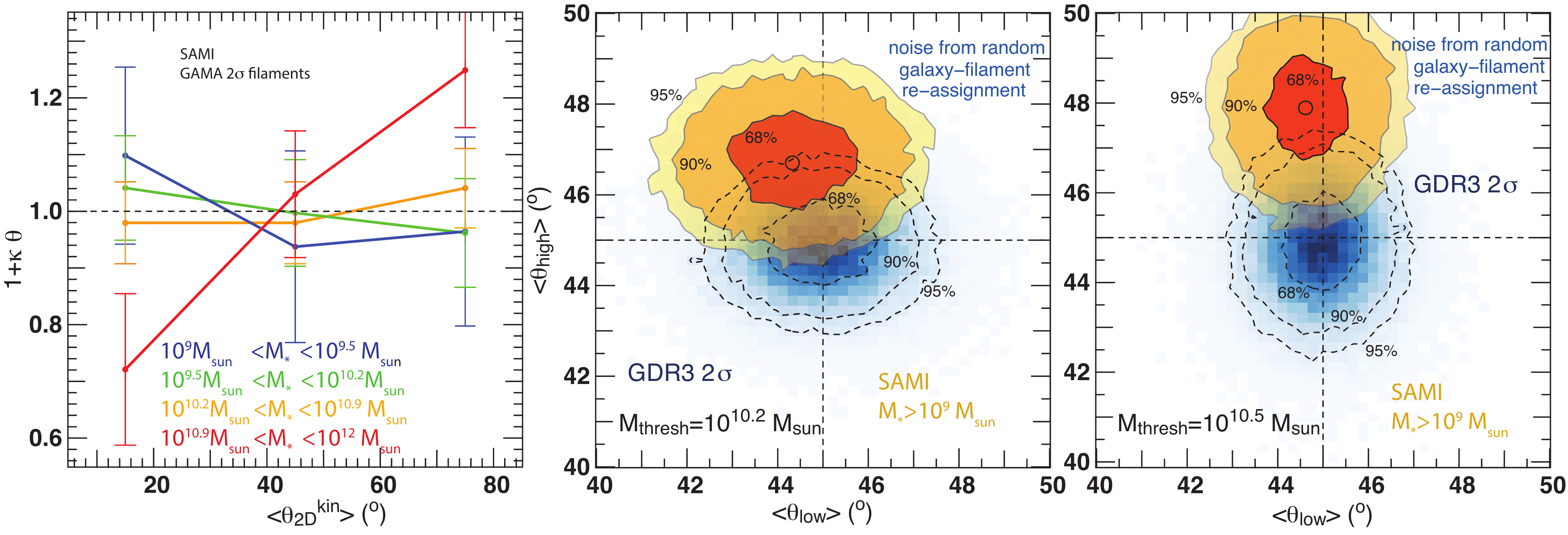}
  \caption{Analysis obtained with the 2$\sigma$ filaments. {\it Left panel}: Similar to Fig.~\ref{fig:align1}. {\it Middle and right panels:} Similar to Fig~\ref{fig:angle2} with mass thresholds $M_{\rm thresh}=10^{10.2}\, M_{\odot}$ and $M_{\rm thresh}=10^{10.5}\, M_{\odot}$. }
\label{fig:2s}
\end{figure*}

Once can see that while large highly contrasted filaments are consistent across the $3\sigma$ and $2\sigma$ reconstruction, the $2\sigma$ network shows also a number of additional finer, lower scale filaments. It is also noisier. 

Fig.~\ref{fig:2s}, left panel, reproduces the analysis of Fig.~\ref{fig:align1} but with the $2\sigma$ reconstruction. It shows the renormalized PDF of $1+\kappa\xi$ of $ \theta_{\rm kin}^{\rm 2D}$ for the SAMI sub-sample with  $10^{9}\, M_{\odot}<M_{*}< 10^{9.5}\, M_{\odot}$ (in blue), with $10^{9.5}\, M_{\odot}M_{*}< 10^{10.2}\, M_{\odot}$ (in green), with $10^{10.2}\, M_{\odot}M_{*}< 10^{10.9}\, M_{\odot}$ (in orange) and with $M_{*}> 10^{10.9}\, M_{\odot}$ (in red). The expected signal for uniformly distributed angles is shown as a horizontal dashed black line.

Results are qualitatively similar to those found for the $3\sigma$ skeleton. But the transition seems to start a lower masses. Indeed, the tendency to display a spin orthogonal the nearby filaments is already detected for $M_{*}>10^{10.2}\, M_{\odot}$ (in yellow). This evolution is consistent with the idea that the transition mass varies with the underlying mass of non-linearity hence the scale (or contrast) of the filaments considered. Note that the evolution appears smoother in this case as the typical mass transition is closer to the peak of the SAMI mass function. The transition is therefore better sampled. 
A caveat is that the low mass alignment trend is harder to recover, hence fainter in this case.

Fig.~\ref{fig:2s}, middle and right panels confirm this observation reproducing Fig~\ref{fig:angle2} with the $2\sigma$ reconstruction. They display as a black circle the average angle in the low-mass subsample $\langle \theta_{\rm low} \rangle$ versus the average angle in the high-mass subsample $\langle \theta_{\rm high} \rangle$ calculated using Method 1 for redshift distortion correction (black circle) and using $M_{\rm thresh}=10^{10.2}\, M_{\odot}$ (middle panel) and $M_{\rm thresh}=10^{10.5}\, M_{\odot}$ (right panel). Red-orange contours are computed using bootstrap and assuming gaussian errors on position angles. Darker to lighter shades indicate the regions in which $50 \%$, $68 \%$  and $90 \%$  of such signals lie. 
Vertical and horizontal black dashed lines show the expectations in each sample for uniformly distributed angles ($45^{\rm o}$). 
Blue shaded areas and dashed contours show the distribution of the spurious noise obtained using the first method described above.   

One can see that both signals are recovered even for a mass threshold as low as $M_{\rm thresh}=10^{10.2}\, M_{\odot}$ and reaches the 3$\sigma$ confidence level for masses $M_{\rm thresh}>10^{10.5}\, M_{\odot}$. In conclusion, finer, lower scale filaments around which low-mass galaxies are more likely to distribute decreases the effective transition mass. This suggests a scale dependence of the transition mass already pointed out in \cite{Cautun15}.

\end{document}